\documentclass[aps,a4paper,preprint,superscriptaddress,preprintnumbers,floatfix,nofootinbib,amsmath,amssymb]{revtex4}

\usepackage{amstext,amssymb}
\usepackage{amsmath}
\usepackage{graphicx}
\usepackage[hyperfootnotes=true]{hyperref}
\usepackage{color}
\usepackage{comment}
\usepackage{array,multirow}
\usepackage{slashed} 
\usepackage{multirow}
\newcommand{\be}{\begin{equation}}
\newcommand{\ee}{\end{equation}}
\newcommand{\bea}{\begin{eqnarray}}
\newcommand{\eea}{\end{eqnarray}}
\newcommand{\nn}{\nonumber}

\def\s1{\hat s}
\def\para{\parallel}

\def\U1mt{U(1)_{L_\mu-L_\tau}}

\def\ol{\overline}
\def\nl{\nonumber\\}
\usepackage{subfigure}

\usepackage{slashed}
\global\long\def\d{\partial}

\begin{document}

\title{\bf Dark matter and flavor anomalies in the light of vector-like fermions and scalar leptoquark}
\author{Suchismita Sahoo$^a$}
\email{suchismita8792@gmail.com}
\author{Shivaramakrishna Singirala$^b$}
\email{krishnas542@gmail.com}
\author{Rukmani Mohanta$^b$}
\email{rmsp@uohyd.ernet.in}
\affiliation{$^a$Department of Physics, Central University of Karnataka, Kalaburagi-585367, India\\
$^b$School of Physics,  University of Hyderabad, Hyderabad-500046,  India}


\begin{abstract}
We make a comprehensive study of vector-like fermionic dark matter and flavor anomalies in a simple extension of standard model. The model is added with doublet vector-like fermions of quark and lepton types, and also a $S_1(\bar{\textbf{3}},\textbf{1},1/3)$ scalar leptoquark. An additional lepton type singlet fermion is included, whose admixture with vector-like lepton doublet plays the role of dark matter and is examined in relic density and direct detection perspective. Electroweak precision observables are computed to put constraint on model parameter space. We constrain the new couplings from the branching ratios and angular observables associated with $b \to sll (\nu_l \bar \nu_l)$, $b \to s \gamma$ decays and also from the recent measurement on muon anomalous magnetic moment.  We then estimate the branching ratios of the rare lepton flavor vioalting $B_{(s)}$ decay modes such as $B_{(s)} \to l_i^\mp l_j^\pm$, $B_{(s)} \to (K^{(*)}, \phi) l_i^\mp l_j^\pm$.   
\end{abstract}

\maketitle
\flushbottom
\section{Introduction} 
The well established fundamental theory of particle physics,  Standard Model (SM), failed to explain the matter-antimatter asymmetry, existence of dark matter and the observation of tiny neutrino mass, which provide a clear indication of the presence  of new physics (NP) beyond it. 
 After the phenomenal discovery of the Higgs boson, though the LHC experiment has not directly  observed any new heavy particles beyond the SM, indirect searches through rare decays of bottom hardons have provided several intriguing hints of NP.   In the last few years,  a collection of interesting deviations from the SM have been manifested in various  angular observables associated with flavor changing neutral currect (FCNC) $b \to s l^+ l^-$ \cite{Aaij:2014ora, Aaij:2019wad, Aaij:2017vbb, Abdesselam:2019wac} and flavor changing charge current (FCCC) $b \to c \tau \bar \nu_\tau$ \cite{Huschle:2015rga, Abdesselam:2016cgx, Abdesselam:2016xqt, Hirose:2017dxl, Hirose:2016wfn, Aaij:2017tyk, Aaij:2015yra, Aaij:2017deq, Aaij:2017uff, Lees:2012xj, Lees:2013uzd, HFLAV} decay modes. Most relevant anomalies include the  lepton flavor  universality violating (LFUV) parameters such as $R_K$ ($2.5\sigma$ deviation) \cite{Aaij:2014ora, Aaij:2019wad, Aaij:2021vac, Bobeth:2007dw, Bordone:2016gaq}, $R_{K^*}$ ($2.2\sigma-2.4\sigma$ deviation) \cite{Aaij:2017vbb, Capdevila:2017bsm}, $R_{D^{(*)}}$ ($3.08\sigma$ deviation) \cite{Amhis:2019ckw, Na:2015kha, Fajfer:2012vx, Fajfer:2012jt}, $R_{J/\psi}$ ($2\sigma$ deviation) \cite{Aaij:2017tyk, Wen-Fei:2013uea, Ivanov:2005fd}  in which the hadronic uncertainties cancelled out significantly. 
The precise analysis of these deviations are needed in both the SM and beyond the SM scenarios in order to probe the structure of NP.

The hypothetical color triplet bosonic particle, leptoquark  arises naturally from the unification of quarks and leptons and its existence  can be found in many extended SM theories \cite{Georgi:1974sy, Georgi:1974my, Fritzsch:1974nn, Langacker:1980js, Pati:1974yy, Pati:1973uk, Pati:1973rp, Shanker:1981mj, Shanker:1982nd, Gripaios:2009dq, Schrempp:1984nj, Kaplan:1991dc}. The scalar (spin $=0$) or vector (spin $=1$) leptoquarks carry both the lepton and baryon quantum numbers. Scenarios with such particles ease to address the flavor anomalies, already investigated in literature \cite{Alok:2017sui, Becirevic:2017jtw, Hiller:2017bzc, DAmico:2017mtc, Becirevic:2016yqi, Bauer:2015knc, Bordone:2016gaq, Calibbi:2015kma, Freytsis:2015qca, Dumont:2016xpj, Dorsner:2016wpm, Varzielas:2015iva, Dorsner:2011ai,  Davidson:1993qk,  Saha:2010vw,  Mohanta:2013lsa, Sahoo:2015fla, Sahoo:2015pzk, Sahoo:2015qha, Sahoo:2015wya, Kosnik:2012dj, Singirala:2018mio, Chauhan:2017ndd, Becirevic:2018afm, Angelescu:2018tyl, Sahoo:2017lzi, Sahoo:2016nvx, Sahoo:2016edx, Sahoo:2020wnk, Bhol:2021iow, Singirala:2021gok, Duraisamy:2016gsd, Sahoo:2016pet}. In the present context, we use a $S_1(\bar{\textbf{3}},\textbf{1},1/3)$ scalar leptoquark to obtain NP contribution to the above quoted anomalies in flavor sector. On the other hand, vector-like fermions are well motivated in low energy scenarios, for review, please see \cite{Ellis:2014dza} and references therein. We consider vector-like fermions of both quark and lepton type, with SM hypercharge assignment. The neutral component of the vector-like lepton can explain the observed relic density of DM in the Universe \citep{Aghanim:2018eyx}. To avoid $Z$-portal direct detection cross section violating the existing spin-independent limits, a trick of mixing with an vector-like singlet lepton is applied in so called singlet-doublet scenarios. Such models are well contained in literature \cite{Mahbubani:2005pt,Arkani-Hamed:2005zuc,DEramo:2007anh,Enberg:2007rp,Cohen:2011ec,Cheung:2013dua,Restrepo:2015ura,Calibbi:2015nha,Cynolter:2015sua,Bhattacharya:2015qpa,Bhattacharya:2017sml,Bhattacharya:2018fus,Freitas:2015hsa,Barman:2019aku,Barman:2019oda,DuttaBanik:2018emv,Bhattacharya:2018cgx,Barman:2019tuo}.

The paper is structured as follows. Section-II provides the particle content, relevant interaction Lagrangian and also the mixing in neutral vector-like lepton sector.  Section-III is presented with the relic density of DM with contributions from annihilations and co-annihilations in various allowed portals. Further, detection prospects are also addressed. Electroweak precision constraints on the model parameters are discussed in section-IV. Section-V discusses the constraints from the quark sector and also from muon anomalous magnetic moment. The implication of new constrained parameters on the lepton flavor violating $B_{(s)}$ decay modes are presented in section-VI. Brief comments on neutrino mass are given in section-VII. Finally, the conclusive remarks are provided in section-VIII. 

\section{New model with leptoquarks}
We extend SM with vector-like  fermion multiplets, one doublet of quark type ($\psi_q$), one doublet of lepton type ($\psi_\ell$) and also a lepton singlet ($\chi_\ell$). The model  also includes a $(\bar{3},1,1/3)$ scalar leptoquark (SLQ) and  the new particles  are assumed to be odd under a discrete $Z_2$ symmetry. The particle content of the model and their quantum numbers are displayed in Table \ref{mutau_model}.
\begin{table}[htb]
\caption{Fields and their charges in the present model.}
\begin{center}
\begin{tabular}{|c|c|c|c|c|}
	\hline
			& Field	& $ SU(3)_C \times SU(2)_L\times U(1)_Y$	& $Z_2$\\
	\hline
	\hline
	Fermions	& $Q_L \equiv(u, d)^T_L$			& $(\textbf{3},\textbf{2},~ 1/6)$	 & $+$\\
			& $u_R$							& $(\textbf{3},\textbf{1},~ 2/3)$	&  $+$	\\
			& $d_R$							& $(\textbf{3},\textbf{1},~-1/3)$	 & $+$\\
			& $\ell_L \equiv(\nu,~e)^T_L$	& $(\textbf{1},\textbf{2},~  -1/2)$	 & $+$\\
			& $e_R$							& $(\textbf{1},\textbf{1},~  -1)$		 & $+$\\
	\hline	
	Vector-like fermions	& $\psi_q \equiv(\psi_u, \psi_d)^T$			& $(\textbf{3},\textbf{2},~ 1/6)$	 & $-$\\
	&  $\psi_\ell \equiv(\psi_\nu,~\psi_l)^T$ & $(\textbf{1},\textbf{2},~ -1/2)$ &   $-$ \\
	&  $\chi_\ell$ & $(\textbf{1},\textbf{1},~ 0)$ &   $-$ \\
\hline	
	Scalars	& $H$							& $(\textbf{1},\textbf{2},~ 1/2)$	&    $+$\\
			& $S_1$						& $ (\bar{\textbf{3}},\textbf{1},~   1/3)$	& $ -$\\    
			\hline
	\hline
\end{tabular}
\label{mutau_model}
\end{center}
\end{table}

The relevant interaction terms emerging from the Lagrangian are given as
\begin{align}
{\cal L} &= -  y_{\ell}\; \ol{{Q_L}^C} S_1 \epsilon^{ab} \psi_{\ell L} -  y^\prime_{\ell}\; \ol{{d_R}^C} S_1 \chi_{\ell R} - y_{q}\; \ol{{\psi_{qL}}^C} S_1 \epsilon^{ab}\ell_L - y^\prime_{q}\; \ol{{Q_L}^C} S_1^*  \epsilon^{ab} \psi_{q L} - y_D \ol{\psi_\ell} \tilde{H} \chi_\ell + {\rm h.c.} \nl &  - M_q \overline{\psi_q}\psi_q - M_\psi \overline{\psi_\ell}\psi_\ell - M_\chi \overline{\chi_\ell}\chi_\ell \,  + \overline{\psi_\ell} \gamma^\mu \left(i \d_\mu - \frac{g}{2} \boldsymbol{\tau}^a\cdot\bold{W}_\mu^a  +\frac{g^{\prime}}{2}B_\mu \right)\psi_{\ell} +  \, \overline{\chi_\ell} \gamma^\mu \left(i \d_\mu\right)\chi_{\ell}\nl &  +    \, \overline{\psi_q} \gamma^\mu \left(i \d_\mu - \frac{g}{2} \boldsymbol{\tau}^a\cdot\bold{W}_\mu^a  -\frac{g^{\prime}}{6}B_\mu \right)\psi_{q}  + \left| \left(i \d_\mu -\frac{g^{\prime}}{3}B_\mu  \right) S_1\right|^2,
\label{eq:Lag}
\end{align}
and the scalar potential of the model is
\begin{align}
V(H,S_1) &=  \mu^2_H  H^\dagger H + \lambda_H (H^\dagger H)^2 
      +\mu^2_{S} ({S_1}^\dagger {S_1})  +\lambda_{S} ({S_1}^\dagger {S_1})^2 +  \lambda_{HS}(H^\dagger_2 H) ({S_1}^\dagger {S_1}).\nonumber
\label{eq:potential}
\end{align}
\subsection{Neutral fermion mass spectrum}
Due to the presence of Dirac term between the lepton multiplets and Higgs in eq. (\ref{eq:Lag}), the new neutral fermions mix among themselves and the corresponding mixing matrix takes the form
\begin{align}
	M_N
	=
	\begin{pmatrix}
		 M_{\psi_\ell}	& M_D	\\
		 M_D	& M_{\chi_\ell}	\\
	\end{pmatrix}, \quad {\rm where} \quad  M_D = \frac{2y_D v}{\sqrt{2}}.
\end{align}
One can diagonalize the above mass matrix using a unitary matrix $
U_\alpha
	=
	\begin{pmatrix}
		 \cos{\alpha}	& -\sin{\alpha}	\\
		 \sin{\alpha}	& \cos{\alpha}	\\
	\end{pmatrix}
$ as $U_{\alpha}^T M_{N} U_{\alpha} = {\rm{diag}}(M_{{N_1}},M_{{N_2}})$, with $\alpha = \frac{1}{2}\tan^{-1}\left(\frac{2M_D}{M_{\psi_\ell} - M_{\chi_\ell}}\right)$. The relation between flavor and mass eigenstates are given by
\begin{equation}
	\begin{pmatrix}
		 \psi_\nu	\\
		 \chi_\ell	\\
	\end{pmatrix} = U_\alpha 	\begin{pmatrix}
		 N_1	\\
		 N_2	\\
	\end{pmatrix}.
\end{equation}
The following relations can be obtained from the above equations:
\begin{eqnarray}
&& M_{\psi_\ell} = M_{N_1} \cos^2 \alpha + M_{N_2} \sin^2 \alpha,\nn\\
&& M_{\chi_\ell} = M_{N_1} \sin^2 \alpha + M_{N_2} \cos^2 \alpha,\nn\\
&& M_D =  \Delta M \sin\alpha\cos\alpha,
\end{eqnarray}
with $\Delta M = (M_{N_1} - M_{N_2})$ representing the mass splitting between  neutral mass eigenstates. The lightest neutral eigenstate ($N_2$) is the DM candidate in the present model. Its annihilations and co-annihilations with other neutral eigenstate ($N_1$) and charged lepton ($\psi_l$) provide relic abundance of the Universe. The mass of charged lepton is $M_{\psi_{\ell}}$ and the mass of both components of $\psi_q$ are equal to $M_{\psi_q}$. In the whole analysis, we consider mass of leptoquark $M_{S_1} = 1.2$ TeV.
\section{Dark sector}
\subsection{Relic abundance}
To compute the freeze-out abundance of vector-like leptonic DM, we utilize the well-known packages, LanHEP \cite{Semenov:1996es} for model implementation and micrOMEGAs \cite{Pukhov:1999gg, Belanger:2006is, Belanger:2008sj} for DM study. The relic density is mainly dictated by three parameters i.e., the mass splitting $\Delta M$, $y_\ell$ and $y^\prime_\ell$. The mass splitting controls the co-annihilation contribution and the Yukawa determines the impact of LQ portal channels on relic density. 

\begin{figure}[thb]
\begin{center}
\includegraphics[width=0.3\linewidth]{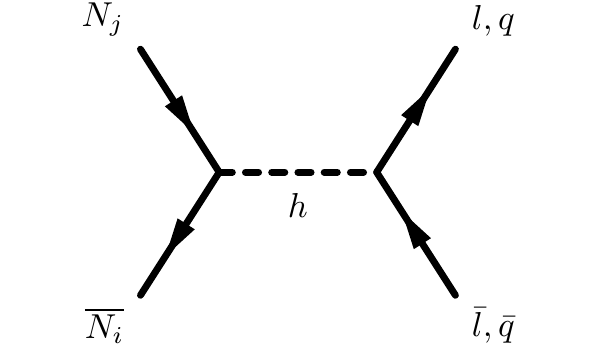}
\includegraphics[width=0.3\linewidth]{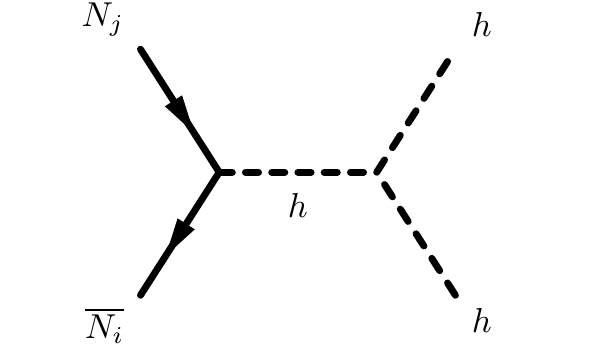}
\includegraphics[width=0.3\linewidth]{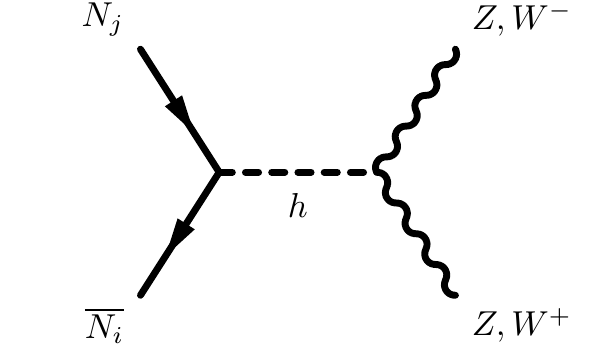}
\includegraphics[width=0.3\linewidth]{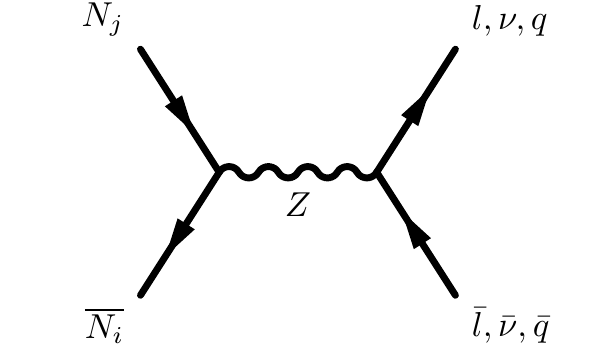}
\includegraphics[width=0.3\linewidth]{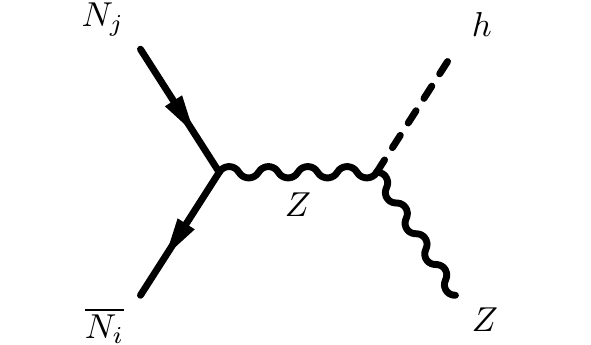}
\includegraphics[width=0.3\linewidth]{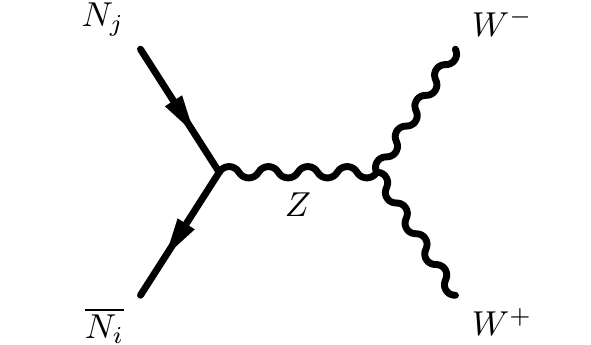}
\includegraphics[width=0.3\linewidth]{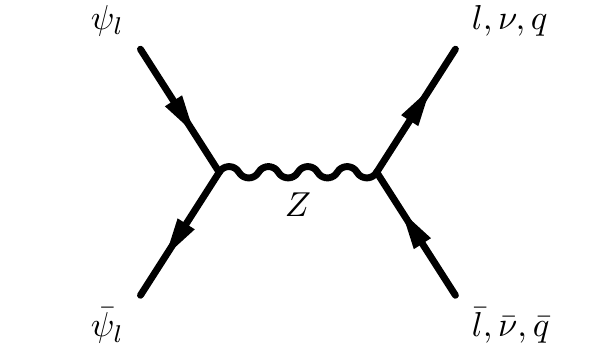}
\includegraphics[width=0.3\linewidth]{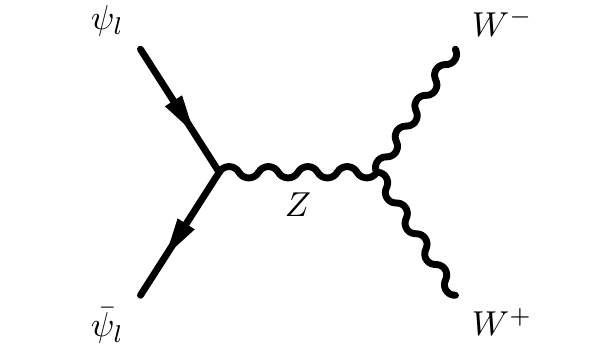}
\includegraphics[width=0.3\linewidth]{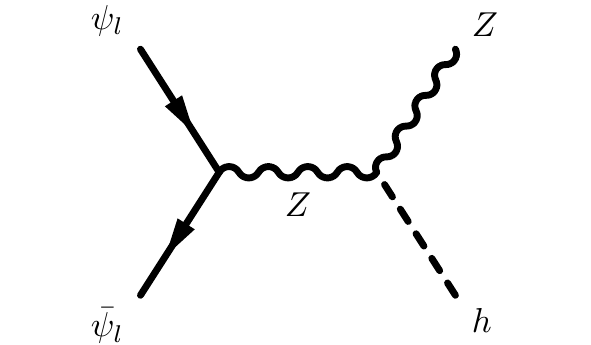}
\includegraphics[width=0.3\linewidth]{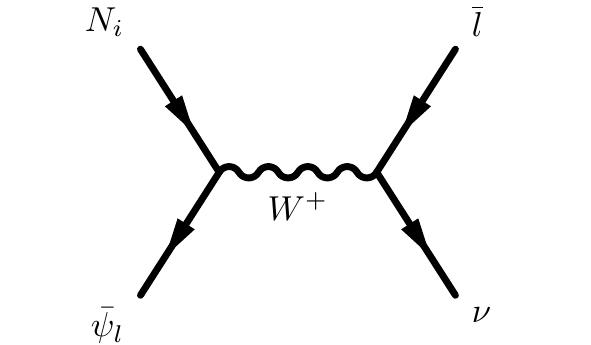}
\includegraphics[width=0.3\linewidth]{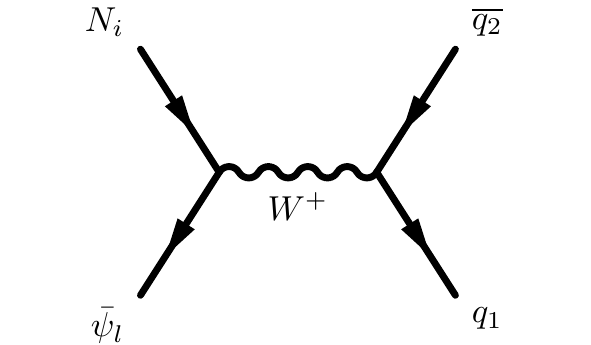}
\includegraphics[width=0.3\linewidth]{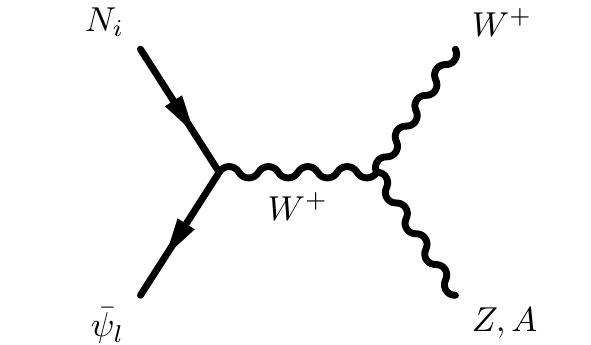}
\includegraphics[width=0.3\linewidth]{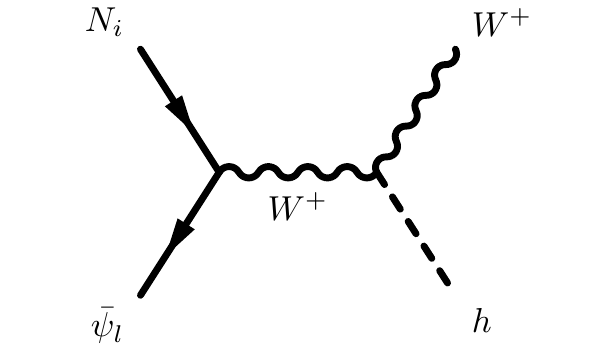}
\includegraphics[width=0.3\linewidth]{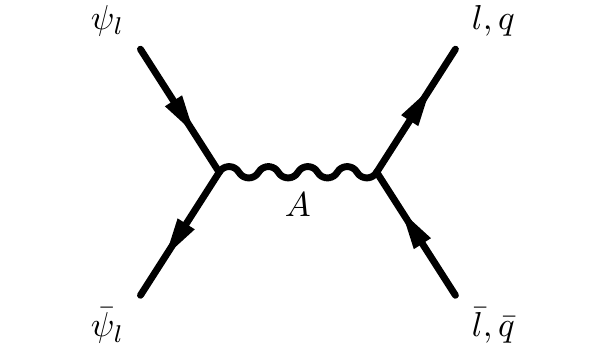}
\includegraphics[width=0.3\linewidth]{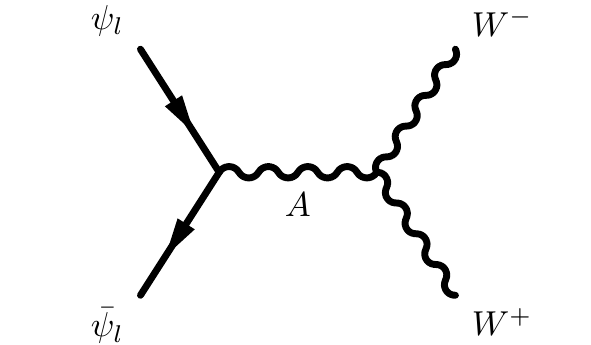}
\caption{Annihilation and co-annihilation channels mediated by SM bosons.}
\label{feyn1}
\end{center}
\end{figure}

\begin{figure}[thb]
\begin{center}
\includegraphics[width=0.3\linewidth]{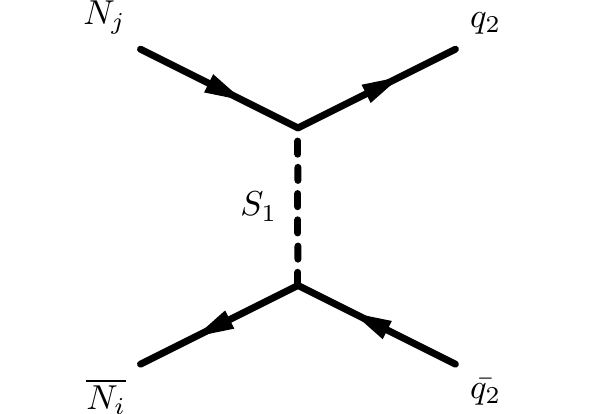}
\includegraphics[width=0.3\linewidth]{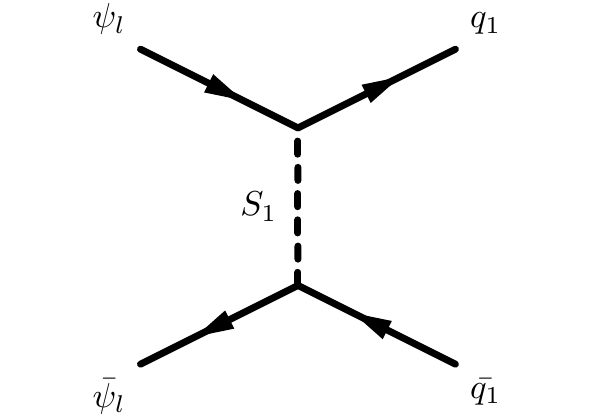}
\includegraphics[width=0.3\linewidth]{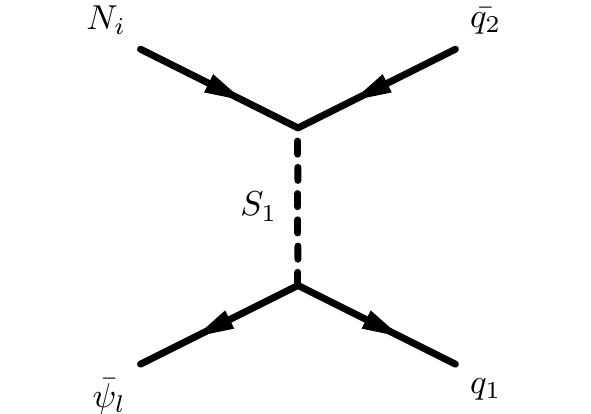}
\includegraphics[width=0.3\linewidth]{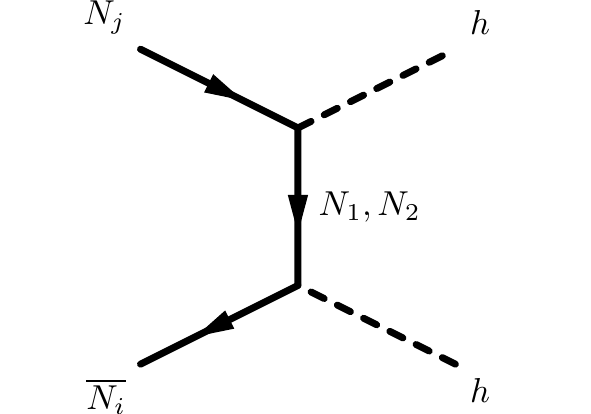}
\includegraphics[width=0.3\linewidth]{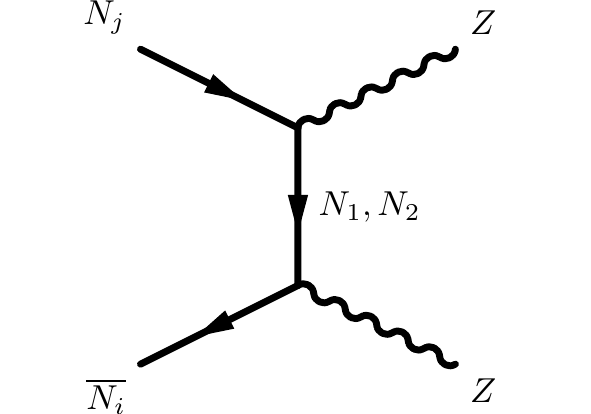}
\includegraphics[width=0.3\linewidth]{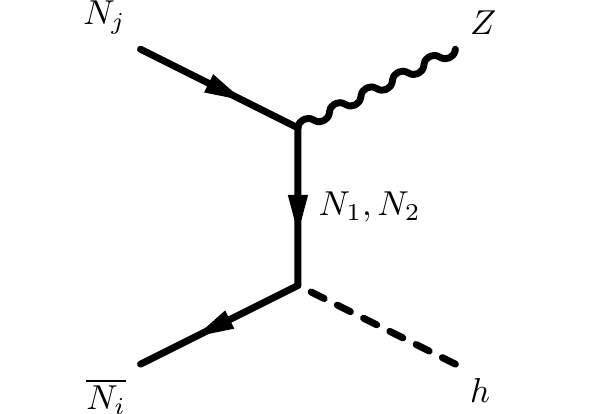}
\includegraphics[width=0.3\linewidth]{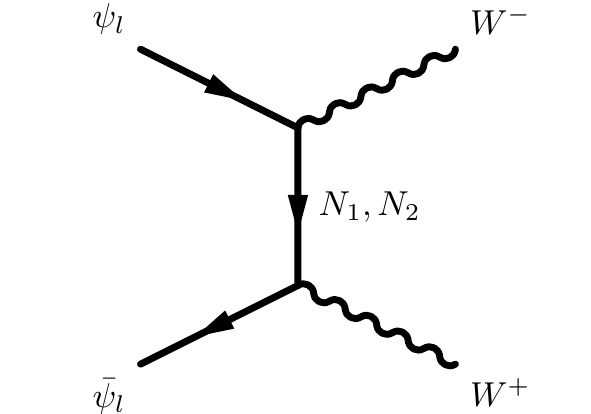}
\includegraphics[width=0.3\linewidth]{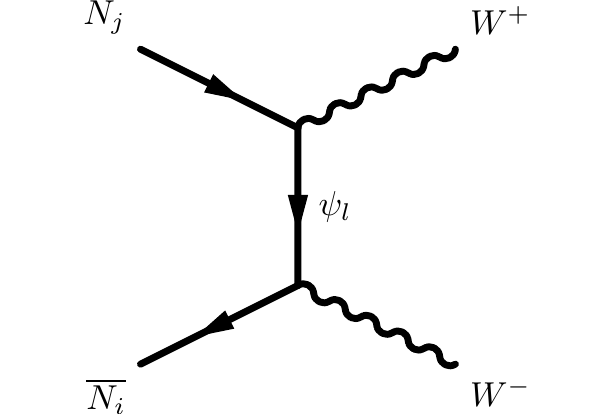}
\includegraphics[width=0.3\linewidth]{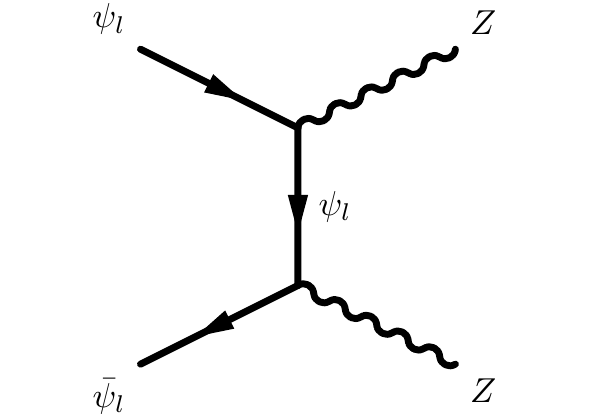}
\includegraphics[width=0.3\linewidth]{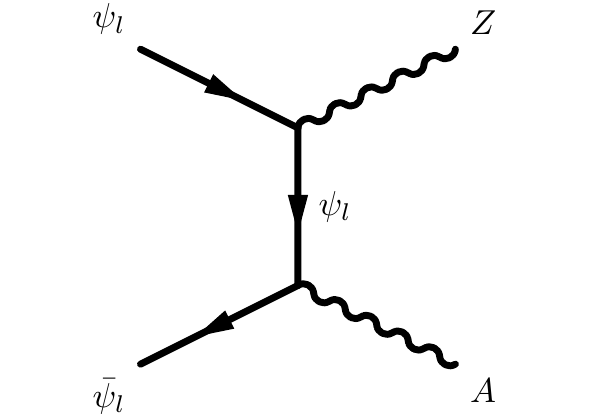}
\includegraphics[width=0.3\linewidth]{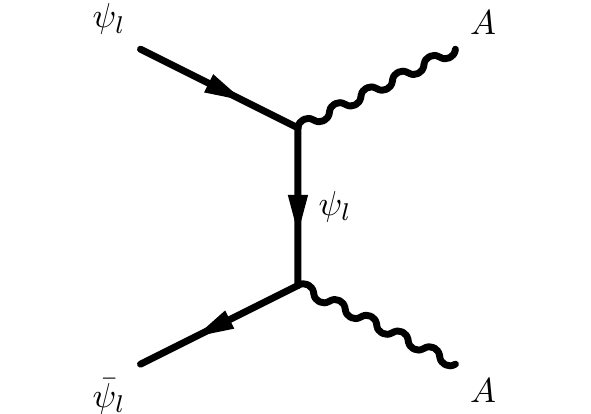}
\includegraphics[width=0.3\linewidth]{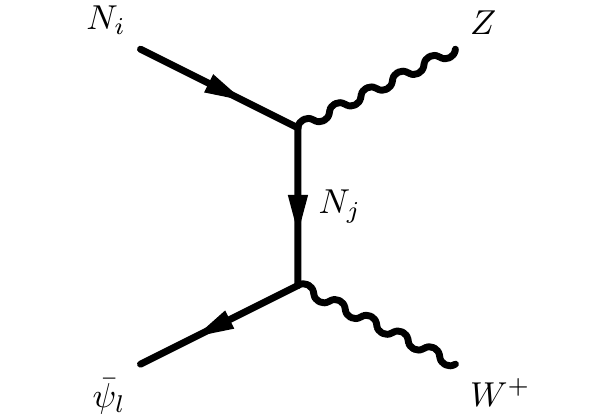}
\includegraphics[width=0.3\linewidth]{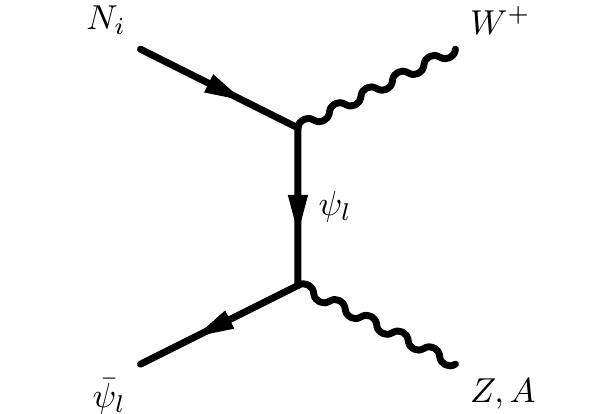}
\includegraphics[width=0.3\linewidth]{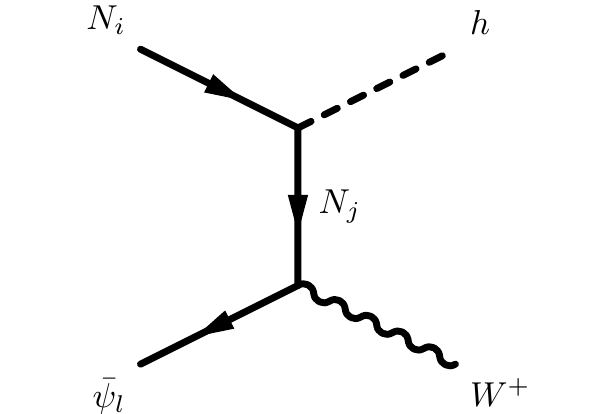}
\caption{Annihilation and co-annihilation channels via new fields portal, where $q_1 = u,c,t$ and $q_2 = d,s,b$.}
\label{feyn2}
\end{center}
\end{figure}

\begin{figure}[thb]
\begin{center}
\includegraphics[width=0.49\linewidth]{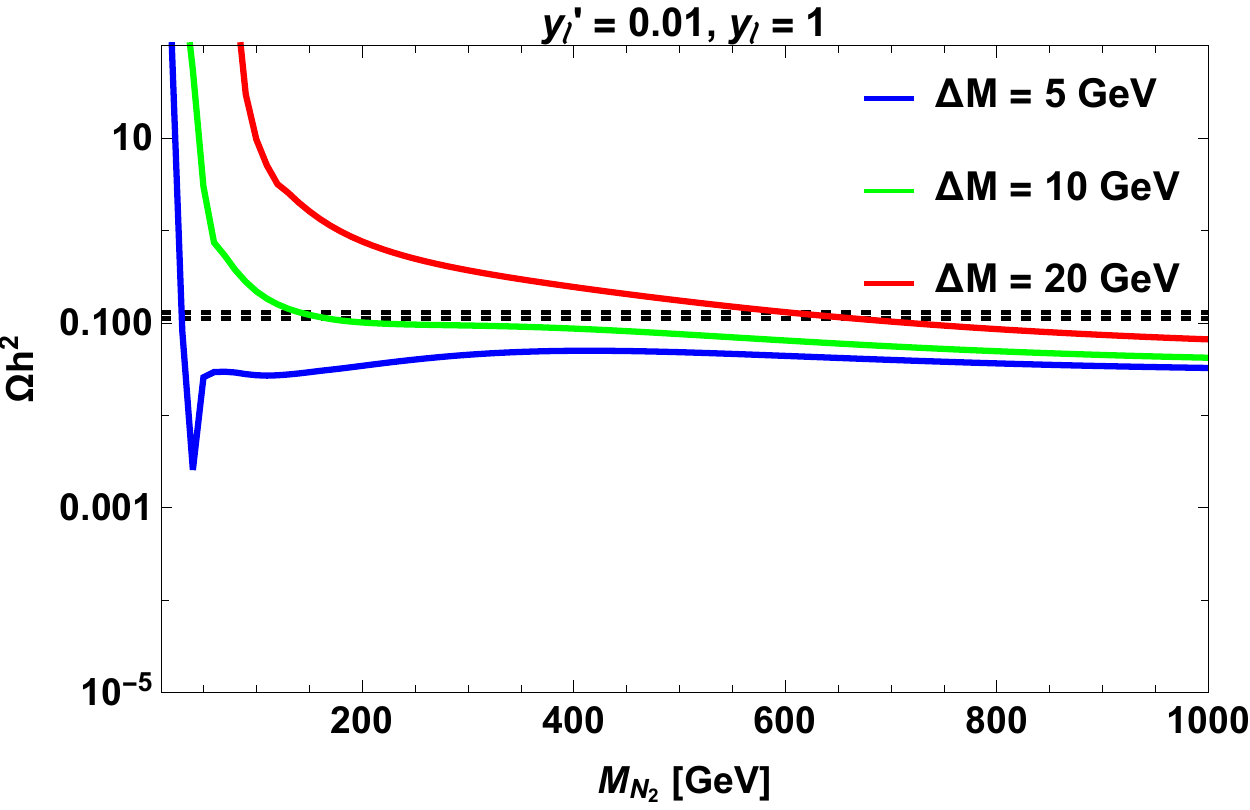}
\includegraphics[width=0.49\linewidth]{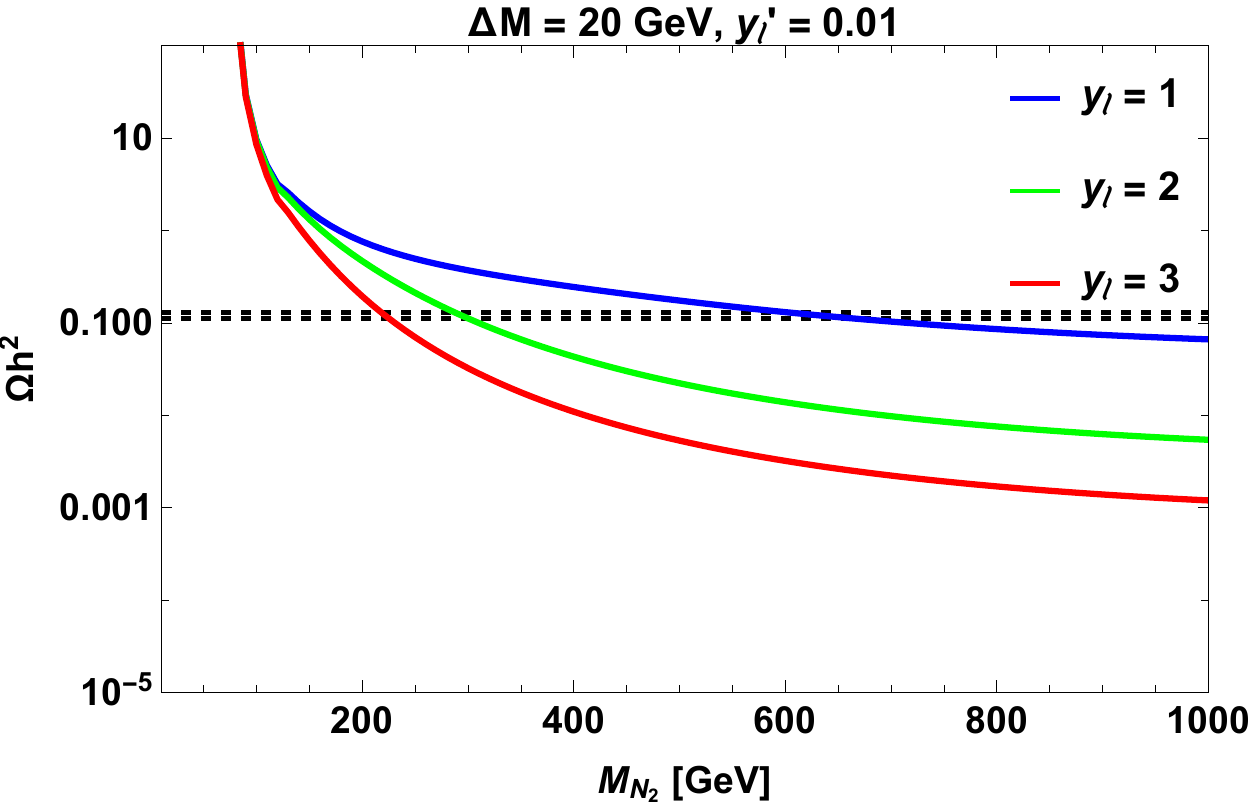}
\caption{Relic density as a function of DM mass. Horizontal dashed lines correspond to Planck limit \cite{Aghanim:2018eyx} in $3\sigma$ region. }
\label{reliccurve}
\end{center}
\end{figure}

All the possible annihilation and co-annihilation channels are displayed in Figs. \ref{feyn1}  and ~\ref{feyn2}. For lower mass splitting, co-annihilation channels of charged and neutral vector-like components (i.e., $N_2, N_1, \psi_l$) can contribute to the total cross section, the impact is made clear in left panel of Fig. \ref{reliccurve}.  The effect of LQ portal channels is directed by the Yukawa $y_\ell$, illustrated in the right panel of Fig. \ref{reliccurve}.  
%
%
\subsection{Direct detection}
\begin{figure}[thb]
\begin{center}
\includegraphics[width=0.3\linewidth]{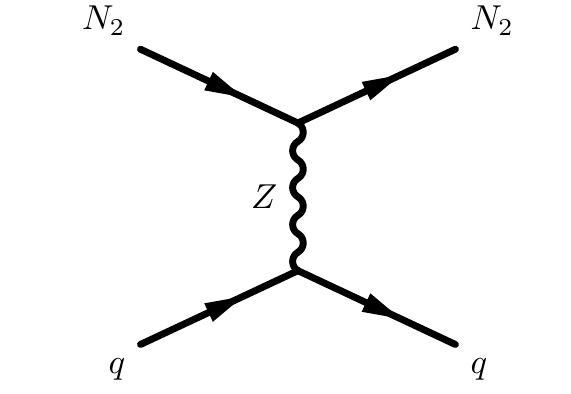}
\includegraphics[width=0.3\linewidth]{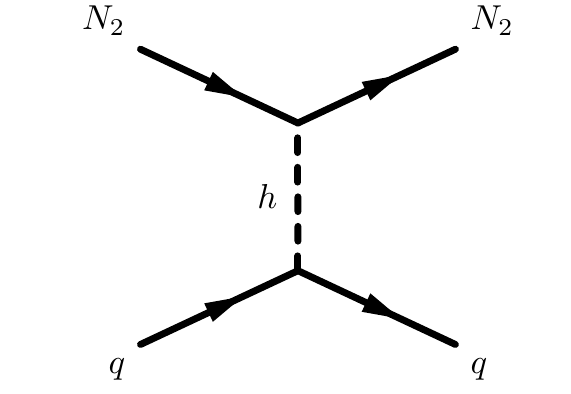}
\includegraphics[width=0.3\linewidth]{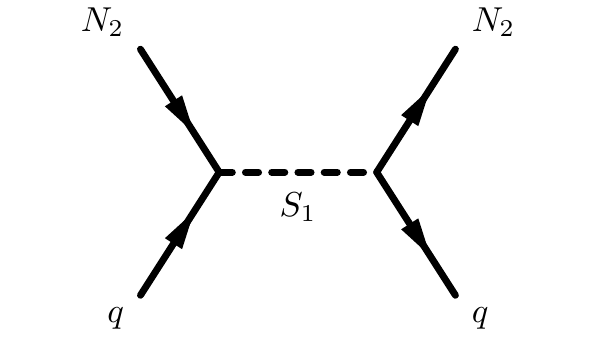}
\caption{Feynman diagrams for direct detection.}
\label{DDfeyn}
\end{center}
\end{figure}

In the present model, the DM can scatter off nucleus in the detector, leaving a footprint through the following effective interactions
\begin{eqnarray}
&&\mathcal{L}^{\rm eff}_{Z} \sim b_q(\overline{q}\gamma^\mu q)(\overline{N_2}\gamma_\mu N_2), \quad  b_q = \frac{\sin^2\alpha}{2M^2_{Z}} (g^\prime \sin\theta_w + g\cos\theta_w) \frac{g C_V^q}{2\cos\theta_w}, \nonumber\\
&&\mathcal{L}^{\rm eff}_{h} \sim a_q (\overline{q} q)(\overline{N_2} N_2), \quad a_q = \frac{\sqrt{2} M_q y_D \sin\alpha \cos\alpha}{M_h^2 v},\nonumber\\
&&\mathcal{L}^{\rm eff}_{S_1} \sim c_q (\overline{q}\gamma^\mu q)(\overline{N_2}\gamma_\mu N_2)+ c^\prime_q (\overline{q}\gamma^\mu \gamma^5 q)(\overline{N_2}\gamma_\mu \gamma_5 N_2), \quad c_q = c^\prime_q = \frac{y^2_{\ell}\sin^2 - y^{\prime 2}_{\ell}\cos^2\alpha}{8 M^2_{S_1}}.
\end{eqnarray}
The interaction is represented as Feynman diagrams in Fig. \ref{DDfeyn} and the corresponding WIMP-nucleon cross sections are given by
\begin{eqnarray}
&&\sigma^{\rm SI}_{Z} = \frac{\mu^2}{\pi} [Z b_p + (A-Z)b_n]^2,\quad b_p = 2b_u + b_d, b_n = b_u + 2b_d,\nn\\
&&\sigma^{\rm SI}_{h} = \frac{\mu^2}{\pi} [Z f_p + (A-Z)f_n]^2,\quad f_{p(n)} = \frac{a_q}{M_q} \left[\frac{2}{9} + \frac{7}{9}(f^{p(n)}_{Tu}+f^{p(n)}_{Td}+f^{p(n)}_{Ts})\right]	,\nn\\
&&\sigma^{\rm SI}_{S_1} = \frac{\mu^2}{\pi} [Z c_p + (A-Z)c_n]^2,\quad c_p = c_d, c_n = 2c_d,\nn\\
&&\sigma^{\rm SD}_{S_1} = \frac{4\mu^2}{\pi} [c^\prime_d \Delta_d + c^\prime_s \Delta_s]^2 J_N(J_N + 1).
\end{eqnarray}
Here $J_N = 1/2$, the typical values of $f^{p(n)}_{Tq}$ and quark spin functions $\Delta_q$ are provided in \cite{Agrawal:2010fh}.

\begin{figure}[thb]
\begin{center}
\includegraphics[width=0.48\linewidth]{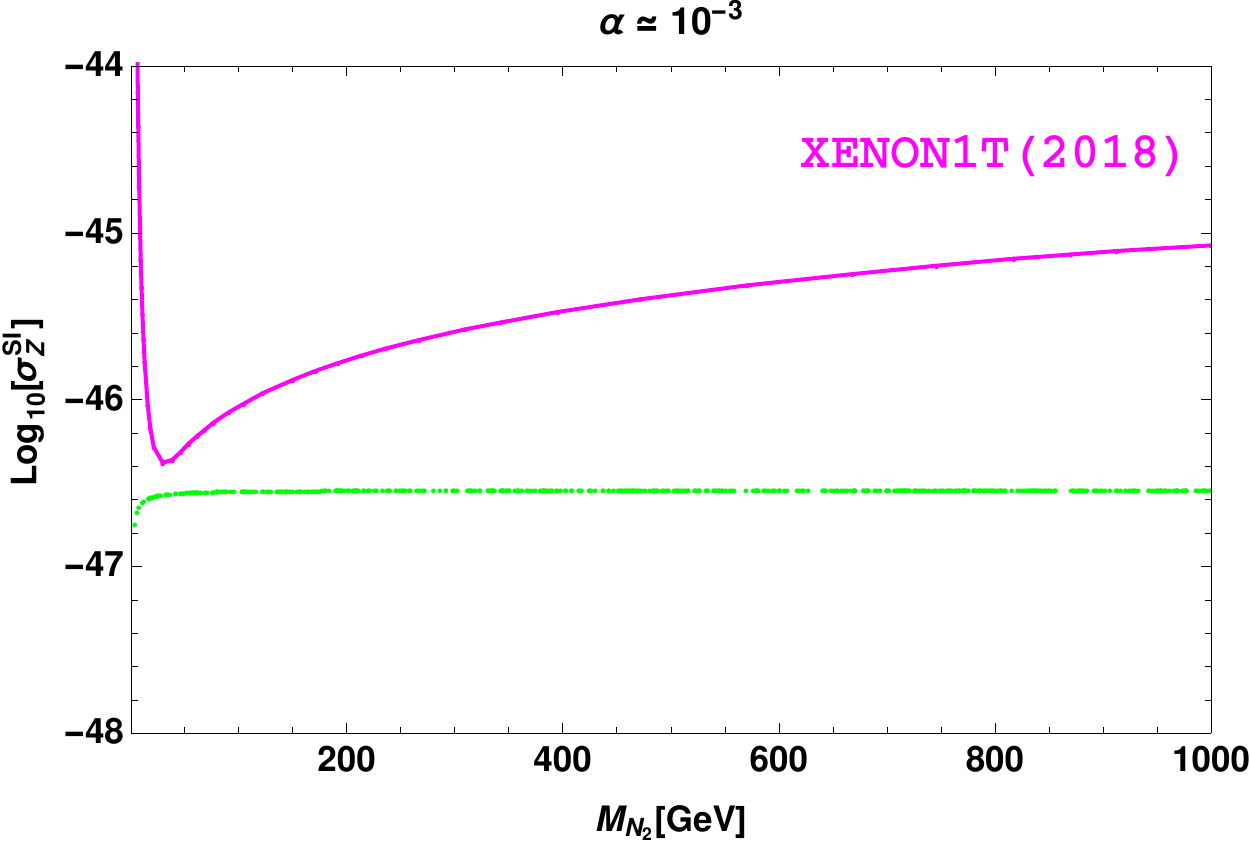}
\includegraphics[width=0.48\linewidth]{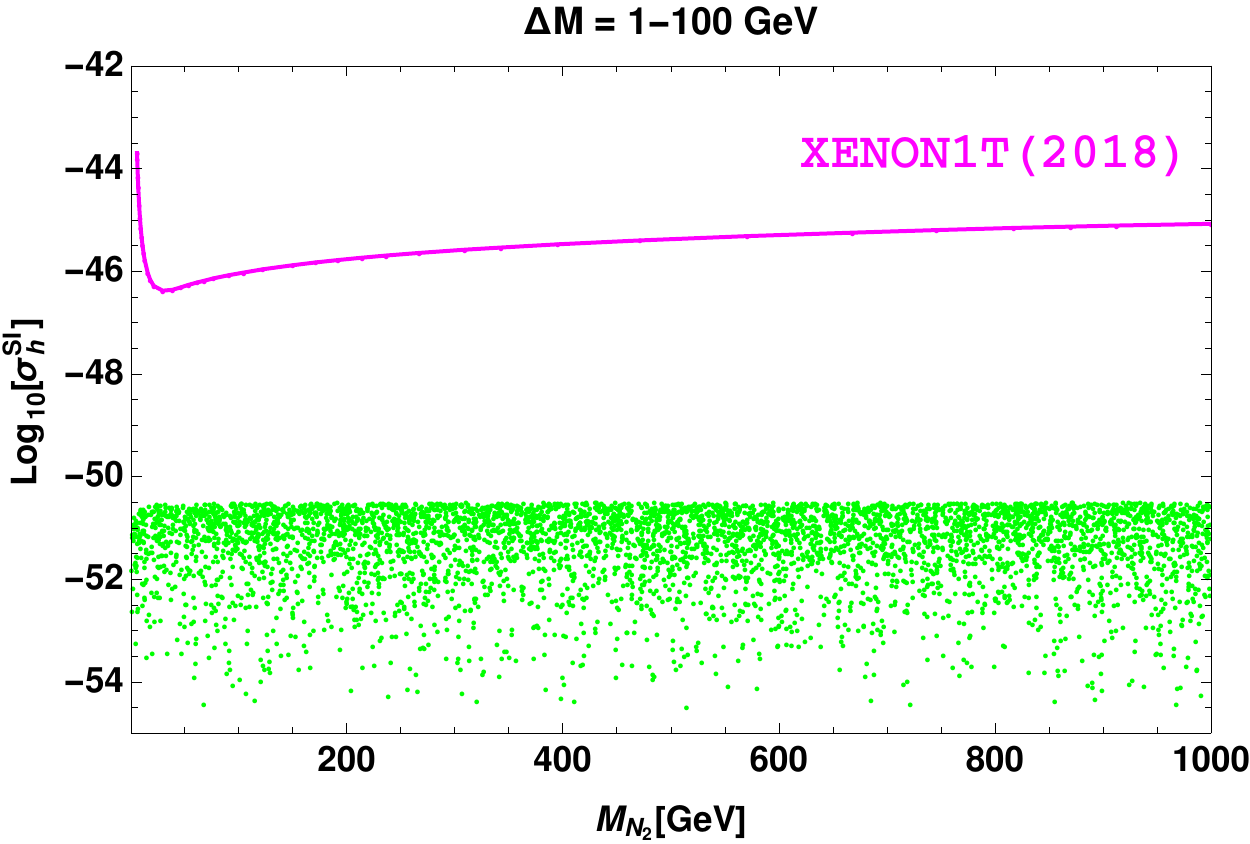}
\hspace{1 cm}
\includegraphics[width=0.48\linewidth]{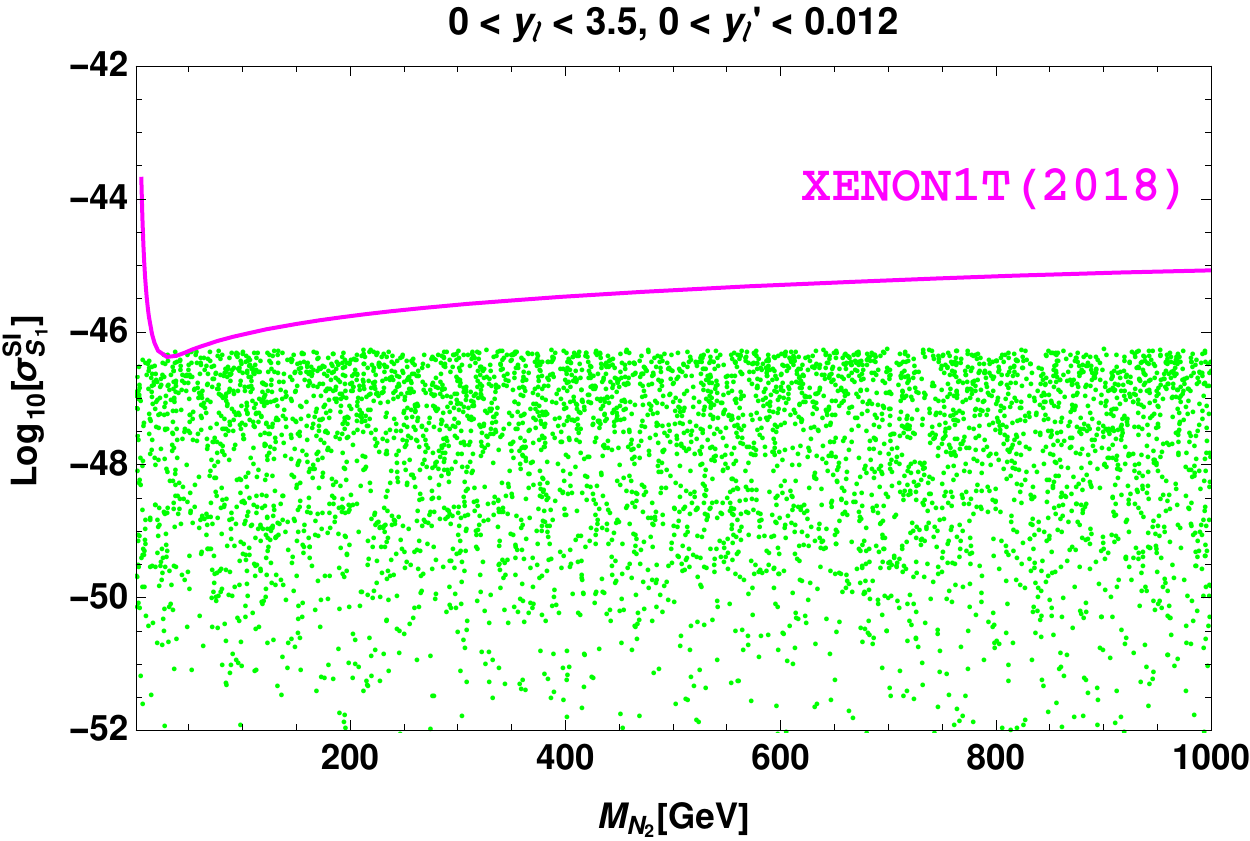}
\includegraphics[width=0.48\linewidth]{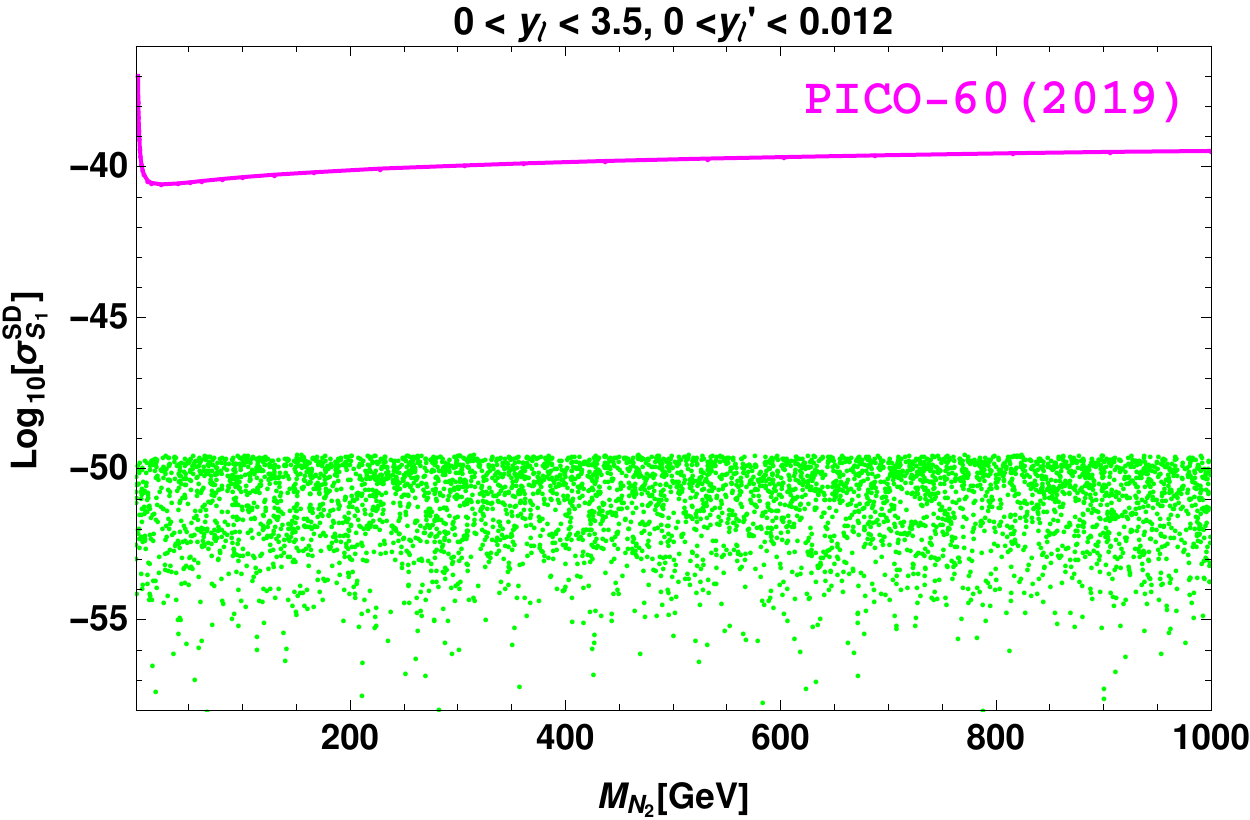}
\caption{WIMP-nucleon cross section in various portals. Magenta lines correspond to upper limits from XENON1T \cite{Aprile:2018dbl} and PICO-60 \cite{Amole:2019fdf}.}
\label{DDcross}
\end{center}
\end{figure}

The smallness of singlet-doublet mixing plays a crucial role in getting the $Z$-portal spin-independent (SI) cross-section within experimental bound of XENON1T \cite{Aprile:2018dbl}. In other words, an upper limit on mixing parameter $\alpha$ ($\sim 10^{-3}$) is levied, as shown in the left plot (upper panel) of Fig. \ref{DDcross}. The same logic is applicable in the context of SLQ portal vectorial part, allowing large Yukawa $y_\ell$ with SI contribution still consistent with XENON1T limit, as displayed in the left plot (lower panel) of Fig. \ref{DDcross}. Right plot in the upper panel corresponds to Higgs-mediated SI contribution, which depends on the mass splitting $\Delta M$. Lower right panel projects SD contribution of axial vector part in SLQ-portal. Above figures suggest the model is safe from the stringent uppper limits of XENON1T \cite{Aprile:2018dbl} and PICO-60 \cite{Amole:2019fdf}.
\section{Electroweak precision parameters}
\begin{figure}[thb]
\begin{center}
\includegraphics[width=0.52\linewidth]{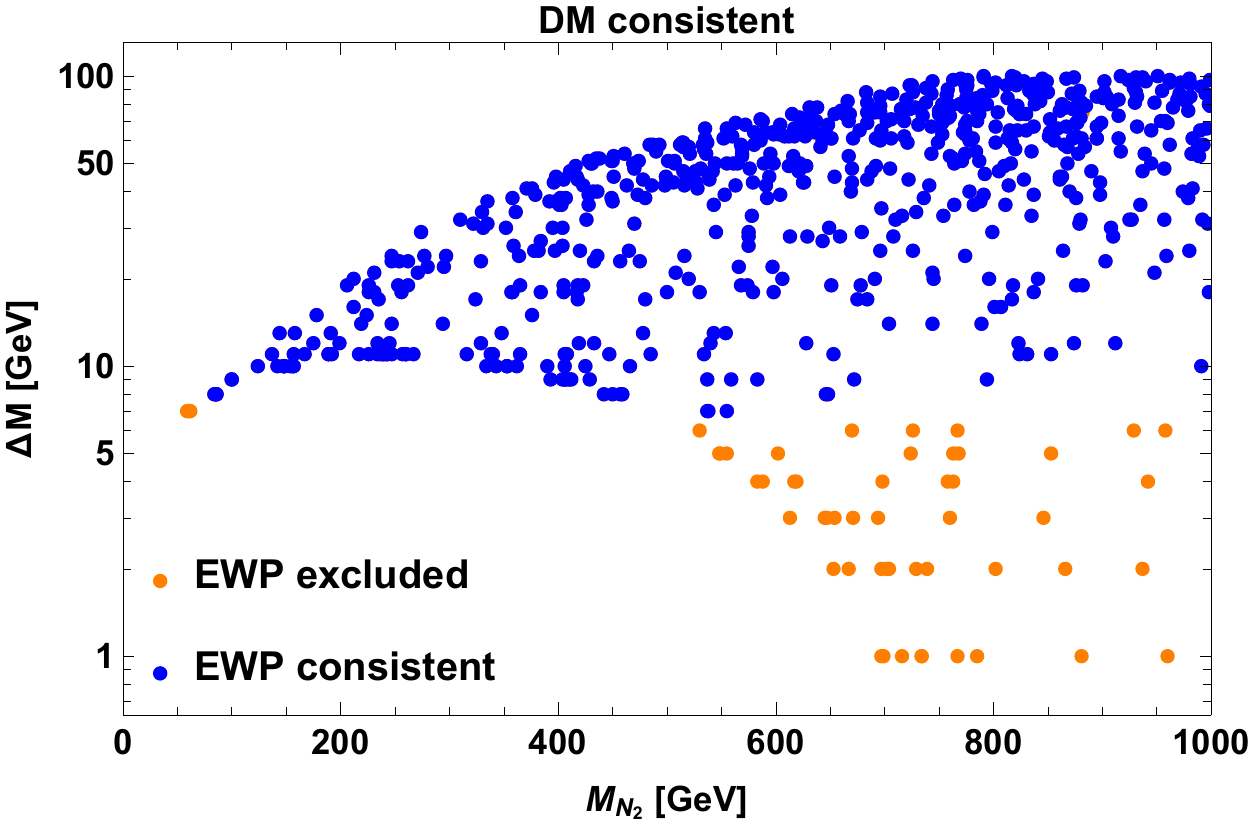}
\caption{Model parameter space with DM and EWP constraints.}
\label{DMEWPscan}
\end{center}
\end{figure}

The vector-like fermions in the present  model can alter the vacuum polarization of SM gauge bosons. The relevant interaction Lagrangian terms read as
\begin{eqnarray}
\mathcal{L}_\ell &\supset & \left(\frac{g}{2} W^3_\mu - \frac{g^\prime}{2} B_\mu\right) \left(\overline{\psi_\nu} \gamma^\mu \psi_\nu\right) +  \frac{g}{\sqrt{2}}W^+_\mu
\left(\overline{\psi_\nu} \gamma^\mu \psi_l\right) + \frac{g}{\sqrt{2}}W^-_\mu
\left(\overline{\psi_l} \gamma^\mu \psi_\nu\right) \nn\\
&&- \left(\frac{g}{2} W^3_\mu + \frac{g^\prime}{2} B_\mu\right) \left(\overline{\psi_l} \gamma^\mu \psi_l\right) ,\\
\mathcal{L}_q &\supset &  \left(\frac{g}{2} W^3_\mu + \frac{g^\prime}{6} B_\mu\right) \left(\overline{\psi_u} \gamma^\mu \psi_u\right) +  \frac{g}{\sqrt{2}}W^+_\mu
\left(\overline{\psi_u} \gamma^\mu \psi_d\right) + \frac{g}{\sqrt{2}}W^-_\mu
\left(\overline{\psi_d} \gamma^\mu \psi_u\right) \nn\\
&&- \left(\frac{g}{2} W^3_\mu - \frac{g^\prime}{6} B_\mu\right) \left(\overline{\psi_d} \gamma^\mu \psi_d\right).
\end{eqnarray}
For the above gauge interactions, the electroweak precision (EWP) parameters are given by \cite{Cynolter:2008ea}
\begin{eqnarray}
&&\hat{S}  =  \frac{g}{g^\prime} \times \Pi'_{W_{3}B}(0),\nn\\
&&\hat{T}  = \frac{1}{M_W^2} \left(\Pi_{W_{3}W_{3}}(0)-\Pi_{W^{+}W^{-}}(0)\right),\nn\\
&&Y  = \frac{M_W^2}{2} \times \Pi''_{BB}(0),\nn\\
&&W  = \frac{M_W^2}{2} \times \Pi''_{W_{3}W_{3}}(0).
\end{eqnarray}
The details of $\Pi$ functions are provided in Appendix~\ref{EWP_apdx}. In the context of vector-like leptons, the mass splitting of neutral components ($\Delta M$) dictate the above parameters. In Fig. \ref{DMEWPscan}, we display the region of parameters which are DM consistent i.e., Planck $3\sigma$ bound on relic density and also limits on WIMP-nucleon cross section from XENON1T and PICO-60. Using the limits of global fit on the precision parameters \cite{Barbieri:2004qk}, the favorable region in the context of precision measurements is also presented. $Y$ and $W$ parameters seem to disfavor low mass splittings. No relevant constraint on the masses of vector-like quarks is obtained from electroweak precision parameters. 

%
%
\section{Constraints from the quark sector}
In this section, we explain the flavor anomalies through one loop box diagrams.   The rare FCNC $b \to s$ transitions  can proceed by the generic  box diagrams in the presence of an additional scalar LQ and the vector-like fermion doublets. The relevant new parameters governing these transitions are the Yukawa-like couplings ($y_{\ell}^{(\prime)}, y_{q}$) and the masses of new particles $(M_{\psi_q}, M_{\psi_l}, M_{N_1}, M_{N_2})$.  The  available new parameter space consistent with the experimental limits of the DM observables are discussed in the previous section. Nevertheless,  the quark  sector can  further constrain these parameters, which will be presented in the subsequent sections.
\subsection{$b \to s ll$ }
\begin{figure}[htb]
\centering
\includegraphics[width=0.85\linewidth]{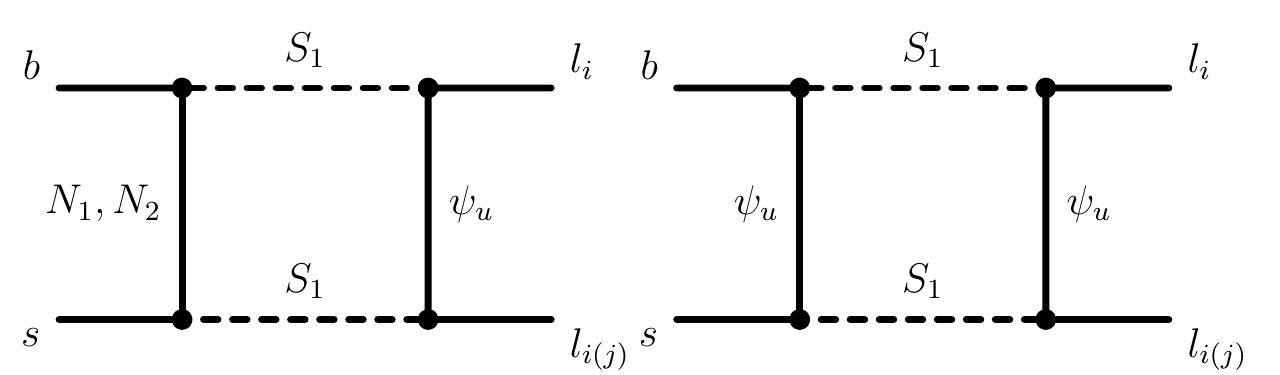}
\caption{One-loop box diagrams of  $b \to s l_il_{i(j)}$ processes  with scalar leptoquark and vector-like fermions in the loop. }\label{Fig:bsll}
\end{figure}

The effective Hamiltonian describing the $ b \to s l^+ l^-$ quark level transition  is given by \cite{Bobeth:1999mk, Bobeth:2001jm} 
\bea
{\cal H}_{\rm eff} &=& - \frac{ 4 G_F}{\sqrt 2} \lambda_t \Bigg[\sum_{i=1}^6 C_i(\mu) \mathcal{O}_i +\sum_{i=7,9,10} \left( C_i(\mu) \mathcal{O}_i+C_i^\prime(\mu) \mathcal{O}_i^\prime\right)\Bigg]\;,\label{ham}
\eea
where  $C_i$'s are  the Wilson coefficients \cite{Hou:2014dza} and ${\cal O}_i$'s are the corresponding four-fermion operators, given as 
\bea
O_7^{(\prime)} &=&\frac{e}{16 \pi^2} \Big(\bar s \sigma_{\mu \nu}
\left (m_s P_{L(R)} + m_b P_{R(L)} \right ) b\Big) F^{\mu \nu}, \nn\\
O_9^{(\prime)}&=& \frac{\alpha_{\rm em}}{4 \pi} (\bar s \gamma^\mu P_{L(R)} b)(\bar l \gamma_\mu l)\;,~~~~~~~ O_{10}^{(\prime)}= \frac{\alpha_{\rm em}}{4 \pi} (\bar s \gamma^\mu 
P_{L(R)} b)(\bar l \gamma_\mu \gamma_5 l)\;, 
\eea
with  $\alpha_{\rm em}$ as  the fine-structure constant and $P_{L,R} = (1\mp \gamma_5)/2$ are the chiral operators. In the SM, though the primed Wilson coefficients $(C_i^\prime)$ are zero, but they can have non-vanishing values in the NP models beyond the SM.    In the presence of scalar LQ, the one loop box diagram responsible for the rare decay processes involving  $b \to s ll$ quark level transition is depicted in Fig. \ref{Fig:bsll}\,. 

  Due to the exchange of leptoquark and new vector like fermions, we obtain additional  contribution to the SM amplitude and the new  Wilson coefficients are given as
\bea \label{C9-C10NP}
C_9^{\rm NP}=-C_{10}^{\rm NP}=&&-\frac{\sqrt{2}|y_{q}|^2}{512\pi G_F \alpha_{\rm em}\lambda_t M_{S_1}^2} \times \Big [|y_\ell \cos \alpha-y_\ell^\prime \sin \alpha |^2 F(x_u,x_{N_1})\nn \\ && +  |y_\ell \sin \alpha +y_\ell^\prime \cos \alpha |^2 F(x_u,x_{N_2})+|y_q^\prime|^2 F(x_u,x_u) \Big]\,,~~~
\eea 
where $x_{N_1}=M_{N_1}^2\cos^2\alpha/M_{S_1}^2$, $x_{N_2}=M_{N_2}^2\sin^2\alpha/M_{S_1}^2$, $x_{u}=M_{\psi_{u}}^2/M_{S_1}^2$ and the loop function has the form
\bea \label{loop-Bsmix}
F(x_i,x_j)&=&\frac{1}{(1-x_i)(1-x_j)}+\frac{x_i^2\log x_i}{(1-x_i)^2(x_i-x_j)}+\frac{x_j^2\log x_j}{(1-x_j)^2(x_j-x_i)}\,.
\eea

 We assume that the NP contribution to $b \to s ee$ transition is negligible i.e., we consider $b \to s ee$ transition is SM like. 
\subsection*{$\boldsymbol{B_s \to l^+ l^-}$} 
 
 In the presence of new Wilson coefficient,  the branching ratio of $B_s \to l^+ l^-$ process is given by 
\bea
{\rm Br}(B_s \to l^+ l^-) = \frac{G_F^2}{16 \pi^3} \tau_{B_s} \alpha^2 f_{B_s}^2  M_{B_s} m_{l}^2   |V_{tb} V_{ts}^*|^2
 \sqrt{1- \frac{4 m_l^2}{M_{B_s}^2}}\left |C_{10}^{\rm SM}+C_{10}^{\rm NP} \right |^2\,.
\eea
The experimental limits on branching ratios of  $B_s \to \mu^+ \mu^-(\tau^+\tau^-)$ processes and the corresponding predicted SM values by using the relevant input parameters from \cite{Zyla:2020zbs} are given by 
\bea
&&{\rm Br}(B_s \to \mu \mu)^{\rm SM}= (3.65 \pm 0.23)\times10^{-9}, ~~
{\rm Br}(B_s \to \mu \mu)^{\rm Expt}=(2.7^{+0.6}_{-0.5} )\times10^{-9},  \nonumber\\
&&{\rm Br}(B_s \to \tau \tau)^{\rm SM}  =   (7.73 \pm 0.49) \times 10^{-7},~~~
{\rm Br}(B_s \to \tau \tau)^{\rm Expt}  <  ~6.8 \times 10^{-3}. 
\eea
\subsection*{$\boldsymbol{B \to K ll}$}
The  differential branching ratio of $B \to K ll$  process with respect to $q^2$ is given by \cite{Bobeth:2007dw}
 \bea
 \frac{d {\rm Br}(B \to K ll)}{d q^2 }= \tau_B \frac{G_F^2 \alpha_{\rm em}^2 |V_{tb}V_{ts}^*|^2}{2^8 \pi^5 M_B^3}\sqrt{\lambda(M_B^2, M_K^2, q^2)} \beta_l f_+^2 \Big ( a_l(q^2)+\frac{c_l(q^2)}{3} \Big )\;,
 \eea
 where
 \bea
 a_l(q^2)&=&  q^2|F_P|^2+ \frac{\lambda(M_B^2, M_K^2, q^2)}{4}
 (|F_A|^2+|F_V|^2) \nn \\ &&+ 2 m_l (M_B^2-M_K^2+q^2) {\rm Re}(F_P F_A^*) +4 m_l^2 M_B^2 |F_A|^2 \;,\nn\\
  c_l(q^2)&=& -   \frac{\lambda(M_B^2, M_K^2, q^2)}{4} \beta_l^2 
 \left (|F_A|^2+|F_V|^2\right ),
 \eea
 with
\bea
F_V & =&\frac{2 m_b}{M_B} C_7^{\rm eff}+ C_9^{\rm eff} +  C_9^{ \rm NP}, ~~~~~F_A = C_{10}^{\rm SM}+C_{10}^{\rm NP},\nn\\
F_P&=&  m_l (C_{10}^{\rm SM}+C_{10}^{\rm NP}) \Big[ \frac{M_B^2 -M_K^2}{q^2}\Big(\frac{f_0(q^2)}{f_+(q^2)}-1\Big) -1 \Big]\;,
\eea  
 and  
\bea
\lambda(a,b,c) =a^2+b^2+c^2-2(ab+bc+ca),~~~~~\beta_l = \sqrt{1-4 m_l^2/q^2}\;.
\eea
 By using the  input parameters from  \cite{Zyla:2020zbs, Colangelo:1996ay}, the predicted branching ratios of $B \to K \mu \mu (\tau \tau)$ processes and the corresponding experimental data \cite{Zyla:2020zbs} are given by
\bea
&&{\rm Br}(  B^0 \to  K^0 \mu^+ \mu^-)^{\rm SM} =(1.48\pm 0.12)\times 10^{-7}\,,~~{\rm Br}( B^0 \to  K^0 \mu^+ \mu^-)^{\rm Expt} = (3.39\pm 0.34)\times 10^{-7}\,,\nn\\
&&{\rm Br}(  B^+ \to  K^+ \mu^+ \mu^-)^{\rm SM} =(1.6\pm 0.13)\times 10^{-7}\,,~~{\rm Br}( B^+ \to  K^+ \mu^+ \mu^-)^{\rm Expt} = (4.41\pm 0.22)\times 10^{-7},\nn \\
&&{\rm Br}(B^+ \to K^+ \tau^+ \tau^-) ^{\rm SM} =(1.52\pm 0.121)\times 10^{-7}\,,~~{\rm Br}(B^+ \to K^+ \tau^+ \tau^-) ^{\rm Expt} < 2.25\times 10^{-3}\,.
\eea

\subsection*{$\boldsymbol{B \to K^*  l^+ l^-}$ }

The differential decay rate of ${B} \rightarrow  K^* l^+ l^-$ process with respect to  $q^2$, after integration over all angles; $\theta_l$ (angle between $l^-$ and 
$B$ in the dilepton frame), $\theta_{K^*}$ (angle between $K^-$ and $B$ in the $K^-\pi^+~$ frame) and $\phi$  (angle between the normal of the $K^-\pi^+$ and the dilepton  planes) \cite{Bobeth:2008ij} is given by
 \bea
 \frac{d\Gamma}{dq^2} = \frac{3}{4} \left(J_1 - \frac{J_2}{3}\right), ~~~~~~J_{1,2} = 2J_{1,2}^s + J_{1,2}^c\,,
 \eea
where the $J_{1,2}^{s(c)}$  function in terms of  transversity amplitudes are given in the Appendix B. The transversity amplitudes as a function of new Wilson coefficients are given as \cite{Altmannshofer:2008dz}
\bea
A_{\perp L,R}&=& N\sqrt{2  \lambda(M_{K^*}^2, M_B^2, q^2)}\Big [ \left ( (C_9^{\rm eff}+C_9^{ \rm NP}) \mp ( C_{10}^{\rm SM}+C_{10}^{\rm NP}) \right )
\frac{V(q^2)}{M_B+M_{K^*}} + \frac{2 m_b}{q^2} C_7 T_1(q^2) \Big]\;,\nn\\
A_{\para L,R}&=& -N \sqrt{2} (M_B^2 -M_{K^*}^2)\Big[ \left ( (C_9^{\rm eff}+C_9^{ \rm NP}) \mp ( C_{10}^{\rm SM}+C_{10}^{\rm NP})  \right ) \frac{A_1(q^2)}{M_B - M_{K^*}} + \frac{2 m_b}{q^2}C_7 T_2(q^2) \Big]\;,\nn\\
A_{0 L,R}& =& - \frac{N}{2 M_{K^*} \sqrt{s}} \Big[ \left ( (C_9^{\rm eff}+C_9^{ \rm NP}) \mp ( C_{10}^{\rm SM}+C_{10}^{\rm NP})  \right )\nn\\
&& \times \left ( (M_B^2 -M_{K^*}^2 -q^2)(M_B + M_{K^*}) A_1(q^2)-\lambda(M_{K^*}^2, M_B^2, q^2) \frac{A_2(q^2)}{M_B + M_{K^*}} \right )\Big]\;,\nn\\
A_t&=& 2N \sqrt{\frac{\lambda(M_{K^*}^2, M_B^2, q^2)}{q^2}} (C_{10}^{\rm SM}+C_{10}^{\rm NP})  A_0(q^2),
\eea
where
\bea
N= V_{tb}V_{ts}^* \left [ \frac{G_F^2 \alpha_{\rm em}^2}{3 \cdot 2^{10} \pi^5 M_B^3} q^2 \beta_l \sqrt{\lambda(M_{K^*}^2, M_B^2, q^2)} \right ]^{1/2}\;. 
\eea
With the use of the particle masss, life time of $B$ meson and the $B \to K^*$ form factor from \cite{Zyla:2020zbs, Ball:2004rg}  the predicted  $B \to K^* ll$  branching ratios and the corresponding experimental data \cite{Zyla:2020zbs} are given by  
\bea
&&{\rm Br}(  B^0 \to  K^{* 0} \mu^+ \mu^-)^{\rm SM} =(1.967\pm 0.158)\times 10^{-8}\,,~{\rm Br}( B^0 \to  K^{* 0} \mu^+ \mu^-)^{\rm Expt} = (9.4\pm 0.5)\times 10^{-7}\,,\nn\\
&&{\rm Br}(B^+ \to K^{* +} \mu^+ \mu^-)^{\rm SM} =(1.758\pm 0.141)\times 10^{-8}\,,~{\rm Br}(B^+ \to K^{*+} \mu^+ \mu^-)^{\rm Expt} < 5.9\times 10^{-7}\,.
\eea 

\subsection*{$\boldsymbol{R_{K^{(*)}}}$}

The updated (full Run-I and Run-II LHCb data) value of the $R_K$ lepton non-universality (LNU) parameter in the  $q^2\in [1.1,6]$~${\rm GeV}^2$  region \cite{Aaij:2021vac}
\bea \label{Eqn:RK-Exp-new}
R_K^{\rm LHCb21} \ = \ \frac{{\rm Br}(B^+ \to K^+ \mu^+ \mu^-)}{{\rm Br}(B^+ \to K^+ e^+ e^-)}=\ 0.846^{+0.042+0.013}_{-0.039-0.012}\,,
\eea
provides the disagreement of $3.1\sigma$ from the SM prediction ~\cite{Bobeth:2007dw, Bordone:2016gaq} 
\bea \label{Eqn:RK-SM}
R_K^{\rm SM} \ = \ 1.0003\pm 0.0001\,.
\eea
Analogous measurements by the LHCb Collaboration on $R_{K^*}$ ratio in two low-$q^2$ bins ~\cite{Aaij:2017vbb}
\bea
R_{K^*}^{\rm LHCb}& \ = \ & \begin{cases}0.660^{+0.110}_{-0.070}\pm 0.03 \qquad q^2\in [0.045, 1.1]~{\rm GeV}^2 \, , \\ 
0.69^{+0.11}_{-0.07}\pm 0.05 \qquad q^2\in [1.1,6.0]~{\rm GeV}^2 \, ,
\end{cases}
\eea 
 have  $2.1\sigma$ and $2.5\sigma$ deviations from their corresponding SM values respectively~\cite{Capdevila:2017bsm}
\bea
R_{K^*}^{\rm SM} \ = \  \begin{cases} 0.92\pm 0.02 \qquad q^2\in [0.045, 1.1]~{\rm GeV}^2 \, , \\  
1.00\pm 0.01\qquad q^2\in [1.1,6.0]~{\rm GeV}^2 \, .
\end{cases}
\eea
Though the Belle experiment \cite{Abdesselam:2019lab, Abdesselam:2019wac} has measured the $R_{K^{(*)}}$ parameters but their results have comparatively larger uncertainties. 
\subsection{$b \to s \nu_l \bar \nu_l$ }
The  effective Hamiltonian of lepton flavor conserving $b \to s \nu_i \bar \nu_i$ process is given by \cite{Buras:2014fpa, Sakaki:2013bfa}
\bea
{{\cal H}_{\rm eff}^{\nu\nu} } =  - \frac{{4{G_F}}}{{\sqrt 2 }}\lambda_t\; C_L^{\rm SM} {\mathcal{O}_L}\,,
\eea
where 
\bea
\mathcal{O}_{L} = \frac{\alpha_{\rm em}}{{4\pi }} [\bar s{\gamma ^\mu }{P_{L}}b][{{\bar \nu }_i}{\gamma _\mu }\left( {1 - {\gamma ^5}} \right){\nu _i}]\,,
\eea
is the six dimensional operator, $C_L^{\rm SM}\approx-X(x_t)/\sin^2\theta_W$ is the SM Wilson coefficient calculated using the loop function $X(x_t)$ \cite{Buras:1998raa} and  $\theta_W$ is  the weak mixing angle. Here $C_L^{ij}$ is zero in the SM. 
\begin{figure}[htb]
\centering
\includegraphics[width=0.85\linewidth]{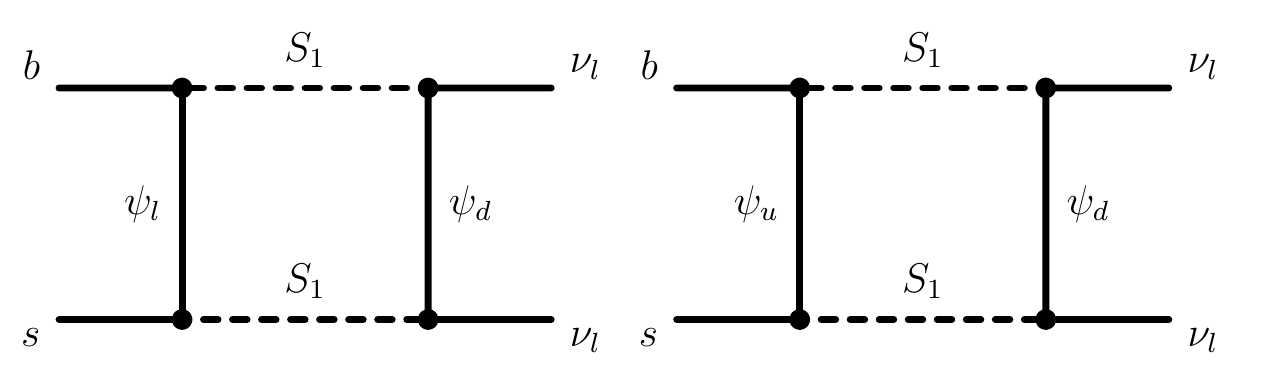}
\caption{One-loop box diagrams of  $b \to s \nu_l \bar \nu_l$ processes  with scalar leptoquark and vector-like fermions in the loop. }\label{Fig:bsnu}
\end{figure}
Fig. \ref{Fig:bsnu} depicts the $b \to s \nu_l \bar \nu_l$ decay mode in the presence of SLQ and vector-like quarks/leptons. 
The additional NP contribution to the SM Wilson coefficient is given by 
\bea \label{CLNP}
C_L^{\rm NP}=&&-\frac{\sqrt{2} |y_{q}|^2}{512\pi G_F \alpha_{\rm em}\lambda_t M_{S_1}^2}\Big [ |y_{\ell}|^2 F(x_d,x_{l}) + |y_q^\prime|^2 F(x_d,x_u) \Big ]\,,
\eea 
where $x_i=M_i^2/M_{S_1}^2$ with $i=\psi_d,~\psi_u,~\psi_l$. 
\subsection*{$\boldsymbol{B \to K \nu_l \bar \nu_l}$}

The branching ratio of $B \to K \nu_l \bar \nu_l$ decay process in the presence of new Wilson coefficient is given by  \cite{Altmannshofer:2009ma}
\bea
\frac{d{\rm Br}}{ds_B} = \tau_B \frac{G_F^2\alpha^2}{256\pi^5} |V_{ts}^* V_{tb}|^2 M_B^5 \lambda^{3/2}(s_B,\tilde{M}_K^2,1) |f_+^K(s_B)|^2 |C_L^{\rm SM} +C_L^{\rm NP}|^2,
\label{BKnunu} 
\eea
where  $\tilde{M}_i = M_i/M_B,~ s_B = s/M_B^2$. Since the different neutrino flavors in the decays $b \to s \nu_l \bar{\nu}_l$
are not distinguished experimentally, the  decay rate will be multiplied with an extra factor $3$.  By using input values from \cite{Zyla:2020zbs}\,, the predicted branching ratio values of $B^{+(0)} \to K^{+(0)} \nu_l \bar \nu_l $  and the corresponding experimental limits are given by
\bea
&&{\rm Br}(B^0 \to K^0 \nu_l  \bar \nu_l)^{\rm SM}=(4.53\pm 0.267) \times 10^{-6}\,,~{\rm Br}(B^0 \to K^0 \nu_l  \bar \nu_l)^{\rm Expt}\textless 2.6 \times 10^{-5}\,,
\nn\\&&{\rm Br}(B^+ \to K^+ \nu_l  \bar \nu_l)^{\rm SM}=(4.9\pm 0.288) \times 10^{-6}\,,~{\rm Br}(B^+ \to K^+ \nu_l  \bar \nu_l)^{\rm Expt}\textless 1.6 \times 10^{-5}\,.
\eea

\subsection*{$\boldsymbol{B_{(s)} \to K^*(\phi) \nu_l \bar \nu_l}$}

In the presence of new physics, the decay rate of $B \to K^* \nu \bar \nu$ process with respect to the $s_B$ and $\cos\theta$  is given as \cite{Altmannshofer:2009ma, Kim:1999waa}
\bea\label{B-Kstar-decay}
\frac{d^2 \Gamma}{ds_B d \cos \theta} = \frac{3}{4} \frac{d \Gamma_T}{ds_B } \sin^2 \theta + \frac{3}{2} \frac{d \Gamma_L}{ds_B } \cos^2 \theta,
\eea
where the longitudinal and transverse polarization decay rates are
\bea
\frac{d \Gamma_L}{ds_B } = 3 m_B^2 |A_0|^2,  \hspace{2cm} 
\frac{d \Gamma_T}{ds_B } = 3 m_B^2 (|A_\perp|^2  + |A_\parallel|^2),
\eea
with the transversity amplitudes, $A_{0, \perp, \parallel}$ in terms of form factor and new Wilson coefficients  are defined as 
\bea
&&A_0 (s_B) = -\frac{N (C_L^{\rm SM}+C_L^{\rm NP}) }{\tilde{M}_{K^*}\sqrt{s_B} } \left[ (1-\tilde{M}_{K^*}^2 - s_B) (1 + \tilde{M}_{K^*}) A_1 (s_B)
 - \lambda(1,\tilde{M}_{K^*}^2, s_B) \frac{A_2 (s_B)}{1 + \tilde{M}_{K^*}} \right]\,,\nn \\
&&A_\perp(s_B) = 2N\sqrt{2} \lambda^{1/2} (1,\tilde{M}_{K*}^2, s_B) (C_L^{\rm SM}+C_L^{\rm NP})  \frac{V(s_B)}{(1+\tilde{M}_{K^*})}\,,\nn \\
&& A_\parallel (s_B)  = -2N \sqrt{2} (1+\tilde{M}_{K^*}) (C_L^{\rm SM}+C_L^{\rm NP}) A_1 (s_B)\,,
\eea
where
\bea
N = V_{tb} V_{ts}^* \left[ \frac{G_F^2 \alpha^2 M_B^3}{3 \cdot 2^{10} \pi^5 } s_B \lambda^{1/2} (1,\tilde{M}_{K^*}^2, s_B) \right]^{1/2}.
\eea
Using all the required input values from \cite{Zyla:2020zbs, Ball:2004rg},   the branching ratio of $B_{(s)} \to K^*(\phi) \nu \bar \nu$ and their corresponding experimental limits \cite{Zyla:2020zbs} are given by
\bea
&&{\rm Br}(B^0 \to K^{* 0} \nu_l  \bar \nu_l)^{\rm SM}=(9.48\pm 0.752) \times 10^{-6}\,,~{\rm Br}(B^0 \to K^{* 0} \nu_l  \bar \nu_l)^{\rm Expt}\textless 1.8 \times 10^{-5}\,,\nn\\
&&{\rm Br}(B^+ \to K^{* +} \nu_l  \bar \nu_l)^{\rm SM}=(1.03\pm 0.06) \times 10^{-5}\,,~{\rm Br}(B^+ \to K^{* +} \nu_l  \bar \nu_l)^{\rm Expt}\textless 4.0 \times 10^{-5}\,,\nn\\
&&{\rm Br}(B_s \to \phi \nu_l  \bar \nu_l)^{\rm SM}=(1.2 \pm 0.07) \times 10^{-5}\,,~{\rm Br}(B_s \to \phi \nu_l  \bar \nu_l)^{\rm Expt}\textless 5.4 \times 10^{-3}\,.
\eea

\subsection{$b \to s \gamma$ }
\begin{figure}[htb]
\centering
\includegraphics[width=0.8\linewidth]{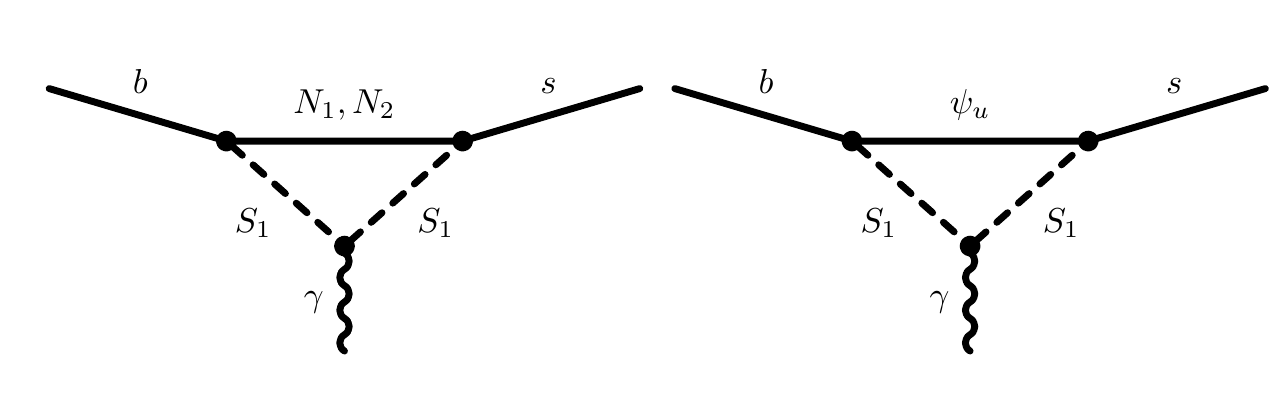}
\caption{One-loop  diagrams of $b \to s \gamma$ processes with scalar  leptoquark and vector-like fermions in the loop. } \label{Fig:bsgamma}
\end{figure}

In Fig.  \ref{Fig:bsgamma}, we present the diagrams of  one loop contributions to  $b \to s \gamma$ processes with scalar leptoquark and vector-like fermions as the internal lines in the loop.
Including the NP contribution, the effective Hamiltonian of  $b\to s\gamma$ decay modes is given by
\bea \label{ham:bsgamma}
  {\cal H}_{\rm eff}^{\gamma}=-\frac{4G_F}{\sqrt{2}}V_{tb}V^*_{ts} (C_7^{\gamma \rm SM}+C_7^{\gamma \rm NP})  \mathcal{O}_7\,,
\eea
where the new $C_7^{\gamma \rm NP}$ Wilson coefficient is given by
\bea \label{c7new}
C_7^{\gamma  \rm NP}=-\frac{\sqrt{2}}{24 G_F V_{tb} V_{ts}^*M_{S_1}^2} &\times & \Big [|y_\ell \cos \alpha -y_\ell^\prime \sin \alpha |^2 \tilde{F_7}(x_{N_1}) + |y_\ell \sin \alpha +y_\ell^\prime \cos \alpha |^2 \tilde{F_7}(x_{N_2})\nn \\ &&+ |y_q^\prime|^2\left( \tilde{F_7}(x_{u}) +2F_7(x_u)\right)\Big ]\,,
\eea
with 
\bea
F_7(x)=\frac{x^3-6x^2+6x\log x+3x+2}{12(x-1)^4},~~~\tilde{F_7}(x)=x^{-1}F_7(x^{-1}).
\eea
\subsection*{$\boldsymbol{\bar B \to X_s \gamma}$}

Including the  SLQ and vector-like fermions contributions,  the total branching ratio  of $B \to X_s \gamma$ decay mode is given by 
\bea \label{tot-bsgamma}
{\rm Br}(\bar B \to X_s \gamma) ={\rm Br}(\bar B \to X_s \gamma)\big |^{\rm SM} \Bigg ( 1+\frac{C_7^{\gamma  \rm NP}}{C_7^{\gamma \rm SM}} \Bigg )^2\,.
\eea
The  SM branching ratio values \cite{Misiak:2015xwa} and the corresponding   experimental limit  \cite{Amhis:2016xyh}  of $B \to X_s \gamma$ decay mode is given by
\bea 
&&{\rm Br}( \bar B \to X_s \gamma)^{\rm SM}_{E_\gamma > 1.6~{\rm GeV}}=(3.36\pm 0.23)\times 10^{-4}\,,\nn\\&&{\rm Br}(\bar B \to X_s \gamma)^{\rm Expt}_{ E_\gamma > 1.6~{\rm GeV}}=(3.32\pm 0.16)\times 10^{-4}\,.
\eea

\subsection{Muon anomalous magnetic moment}
\begin{figure}[htb]
\centering
\includegraphics[width=0.9\linewidth]{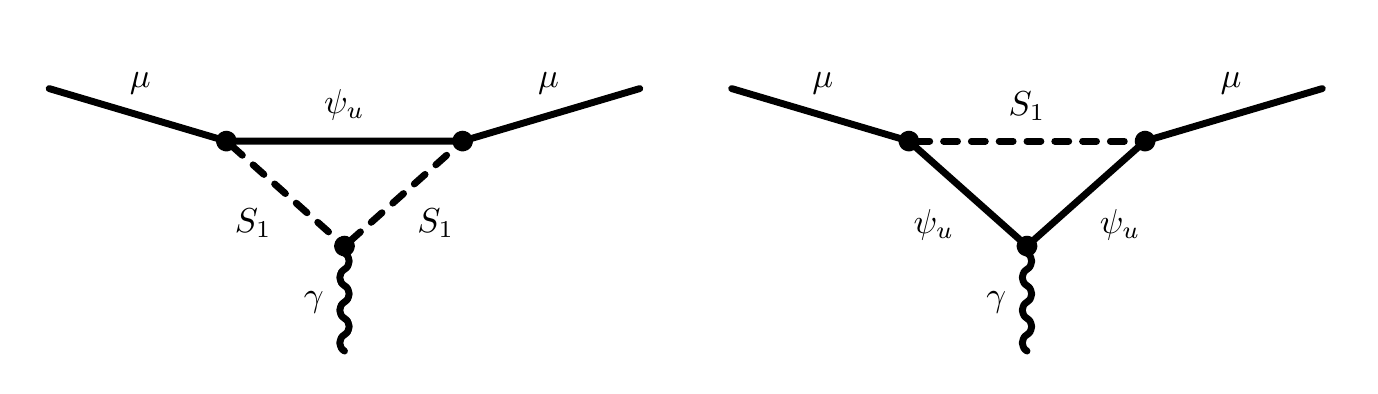}
\caption{One-loop  contribution to muon anomalous magnetic moment with scalar  leptoquark and vector-like fermion in the loop. } \label{Fig:muon}
\end{figure}

 The long sustaining discrepancy between the experimental measurements and SM value of muon anomalous magnetic moment took further step with the recent observations from Fermilab.  Earlier, E$821$ experiment at Brookhaven National laboratory \cite{Bennett:2006fi} has reported a deviation of $3.3\sigma$ from SM prediction \cite{Aoyama:2020ynm}
\begin{equation}
\Delta a^{\rm BNL}_\mu = a^{\rm exp}_\mu - a^{\rm SM}_\mu = (26.1\pm 7.9) \times 10^{-10}.
\end{equation}
With lattice calculations, the deviation mounts to $3.7\sigma$ \cite{Blum:2018mom,Keshavarzi:2018mgv}. Recently, Fermilab's E$989$ experiment \cite{Abi:2021gix} has announced a discrepancy of $4.2\sigma$ 
\begin{equation}
\Delta a^{\rm FNAL}_\mu  = (25.1\pm 5.9) \times 10^{-10}.
\end{equation}
The absolute magnitude of the discrepancy between the SM prediction and the experimental value  is small and can be accommodate by adding the new physics contributions. Fig. \ref{Fig:muon} represents the one loop contribution to the muon anomalous magnetic moment with scalar leptoquark and vector-like quark in the loop. 

  The scalar LQ contribution to $a_\mu$ is 
\bea
\Delta a_\mu &=&  \frac{m_\mu^2 (y_q)^2}{8\pi^2 M_{S_1}^2}  \Big (2 (f_1(x_u)+f_2(x_u))-(\bar{f_1}(x_u)+\bar{f_2}(x_u)) \Big )\,,
\eea
where the functions $f_{1,2}(x_u)$ and $\bar{f}_{1,2}(x_u)$ are defined in Appendix C. 

\begin{figure}[htb]
\centering
\includegraphics[scale=0.48]{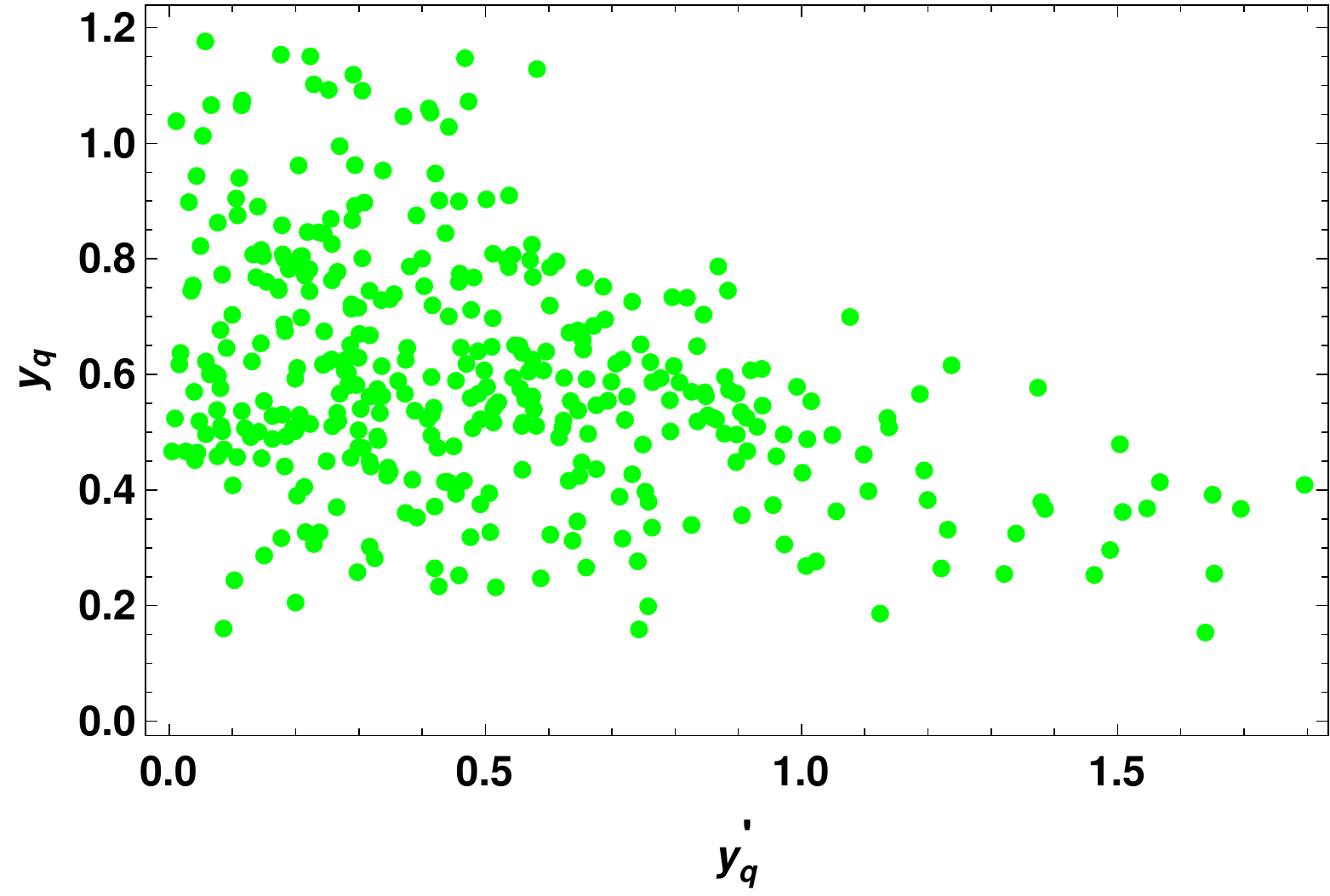}
\quad
\includegraphics[scale=0.435]{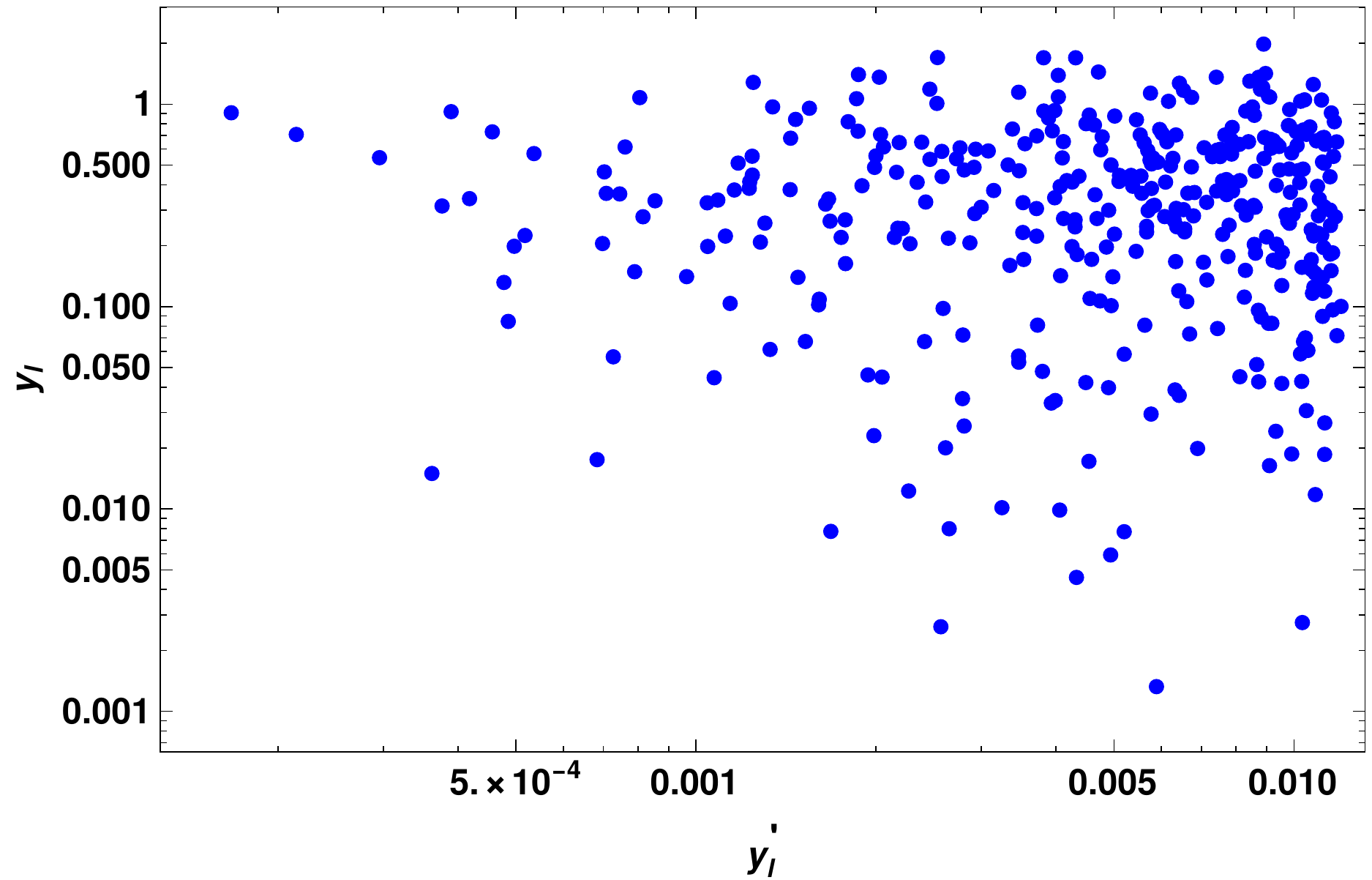}
\caption{Constraints on $y_q^\prime-y_q$ (left panel) and $y_\ell^\prime-y_\ell$ (right panel) parameters obtained from the branching ratios of $B \to K^{(*)} l^+ l^- (\nu_l \bar \nu_l)$, $B_s \to ll$ ($l=\mu,\tau$), $\bar B \to X_s \gamma$ processes, $R_{K^{(*)}}$ ratios and the muon anomalous magnetic moment. }\label{Fig:conc}
\end{figure}

Using all the above discussed observables, the predicted allowed region for  $y_q^\prime-y_q$ (left panel) and $y_\ell^\prime-y_\ell$ (right panel) parameters are presented in Fig. \ref{Fig:conc}\, and the allowed parameter space are given in Table \ref{Tab:conc}\,.

\begin{table}[htb]
\centering
\begin{tabular}{|c|c|c|c|c|c|}
\hline
Parameters ~&~$y_q$~&~$y_q^\prime$~&~$y_\ell$~&~$y_\ell^\prime$~\\
\hline
~Allowed values~&~$0-1.2$~&~$0-1.8$~&~$0-2.0$~&~$0-0.012$~\\
\hline
\end{tabular}
\caption{Allowed parameter space of $y_q^{(\prime)}$ and $y_\ell^{(\prime)}$. }\label{Tab:conc}
\end{table}

\section{Impact on lepton flavor violating $B$ decays  }

After getting constraints on new parameters such as leptoquark couplings and vector-like fermions masses, we now look for its implication on the rare lepton flavor violating (LFV) $B$  decay modes.

The LFV $B$ meson decays mediated by the $b \to s l_i l_j$ quark level transitions are suppressed in the SM. However, it can be studied in the presence of SLQ. 
The one loop box diagram  for $b \to s l_i l_j$ processes mediated via the SLQ and vector-like fermions is presented in  Fig. \ref{Fig:bsll}\,. In the presence of NP, the  effective Hamiltonian for $b \to s l_i^- l_j^+$ processes is given by  
\bea
\mathcal{H}_{\rm eff}\left( b \to s l_i^- l_j^+ \right) \ = -\frac{G_F \alpha_{\rm em}}{\sqrt{2} \pi}  V_{tb} V_{ts}^* \Big[ C_9^{\rm NP} \left( \bar{s} \gamma^\mu P_L b \right) \left(\bar{l}_i \gamma_\mu l_j\right) +C_{10}^{\rm NP} \left( \bar{s} \gamma^\mu P_L b \right) \left(\bar{l}_i \gamma_\mu \gamma_5 l_j\right)\Big]\,,
\label{ham-LFV}
\eea
where $C_{9, 10}^{\rm NP}$ are defined in Eqn. \ref{C9-C10NP}\,.

\subsection{$B_s \to l_i^- l_j^+$}

The branching ratio of the LFV $B_s\to l_i^- l_j^+$ decay process is given as ~\cite{Becirevic:2016zri}
\begin{align}\label{Eq:LFV-Bsll}
& {\rm Br}(B_s\to  l_i^-l_j^+) \ = \ \tau_{B_s} \frac{\alpha_\mathrm{em}^2 G_F^2}{64\pi^3M_{B_s}^3}f_{B_s}^3 |V_{tb}V_{ts}^*|^2 \lambda^{1/2}(M_{B_s},m_i,m_j)\nonumber\\
&\times\Bigg [[M_{B_s}^2-(m_i+m_j)^2]  (m_i-m_j)^2 |C_9^{\rm NP}|^2 +[M_{B_s}^2-(m_i-m_j)^2](m_i+m_j)^2 |C_{10}^{\rm NP}|^2 \Bigg ]\,.
\end{align}

\subsection{$B \to K l_i^- l_j^+$}

The differential branching ratio of $\overline{B}\to \overline{K}l_i^- l_j^+$ process is given as~\cite{Duraisamy:2016gsd}
\bea \label{Eq:LFV-BKll}
\frac{{\rm d}{\rm Br}}{{\rm d}q^2}(\overline{B}\to \overline{K}l_i^- l_j^+) &\ = \ & \tau_B \frac{ G_F^2 \alpha^2_{\rm em} }{2^{12} \pi^5 M_B^3}\beta_{ij} \sqrt{\lambda(M_B^2, M_K^2, q^2)} |V_{tb}V_{ts}^*|^2 \left(J_1+J_2\right)\,,
\eea
where
\bea
J_1 & \ = \ & 4\Bigg[ \left( 1-\frac{(m_i-m_j)^2}{q^2}\right) \frac{1}{3} \left( 2q^2+\left(m_i+m_j\right)^2 \right) |H_V^0|^2  \nn \\  && \quad + \frac{(m_i-m_j)^2}{q^2} \left(q^2-\left(m_i+m_j\right)^2 \right) |H_V^t|^2 \Bigg], \nn \\
J_2 & \ = \ &4\Bigg[ \left( 1-\frac{(m_i+m_j)^2}{q^2}\right) \frac{1}{3} \left( 2q^2+\left(m_i-m_j\right)^2 \right) |H_A^0|^2 \nn \\ && \quad + \frac{(m_i+m_j)^2}{q^2} \left( q^2-\left(m_i-m_j\right)^2 \right) |H_A^t|^2 \Bigg]\,,
\eea
with
\bea
&&H_V^0 \ = \ \sqrt{\frac{\lambda(M_B^2,M_K^2,q^2)}{q^2}} f_+(q^2)C_9^{\rm NP}, ~~~H_V^t \ = \ \frac{M_B^2-M_K^2}{\sqrt{q^2}} f_0(q^2)C_9^{\rm NP}, \nn \\
&&H_A^0 \ = \ \sqrt{\frac{\lambda(M_B^2,M_K^2,q^2)}{q^2}}f_+(q^2)C_{10}^{\rm NP}, ~~~H_A^t \ = \ \frac{M_B^2-M_K^2}{\sqrt{q^2}} f_0(q^2) C_{10}^{\rm NP}\,,
\eea
and 
\bea
\beta_{ij} & \ = \ & \sqrt{\left( 1-\frac{(m_i+m_j)^2}{q^2}\right) \left( 1-\frac{(m_i-m_j)^2}{q^2}\right)}\,.
\eea

\subsection{$B \to K^* l_i^- l_j^+$ and $B_s \to \phi l_i^- l_j^+$}
The differential branching ratio of $\overline{B}\to \overline{K}^* l_i^- l_j^+$ decay process  in the presence of SLQ and vector-like fermions is given by~\cite{Becirevic:2016zri}
\bea\label{Eq:LFV-BKsll}
\frac{\mathrm{d}{\rm Br}}{\mathrm{d}q^2} (\overline{B}\to \overline{K}^*  l_i^- l_j^+) \ = \ \dfrac{1}{4} \left[ 3 I_1^c(q^2)+6 I_1^s(q^2)-I_2^c(q^2)-2 I_2^s(q^2) \right], 
\eea
where the angular coefficients $I_i(q^2)$ are given by~\cite{Nebot:2007bc} 
\begin{align}
I_1^s(q^2) & \ = \ \Big{[} |A_\perp^L|^2 +|A_\parallel|^2+(L\to R)\Big{]} \dfrac{\lambda_q+2[q^4-(m_i^2-m_j^2)^2]}{4 q^4}+\dfrac{4 m_i m_j}{q^2}\mathrm{Re}\left(A_\parallel^L A_\parallel^{R\ast}+A_\perp^L A_\perp^{R\ast}\right),\nonumber \\
I_1^c(q^2) & \ = \ \left[ |A_0^L|^2+|A_0^R|^2\right]\dfrac{q^4-(m_i^2-m_j^2)^2}{q^4}+\dfrac{8 m_i m_j}{q^2} \mathrm{Re}\left( A_0^L A_0^{R\ast} - A_t^L A_t^{R\ast}\right)\nonumber\\
&\qquad \qquad\qquad\qquad-2\dfrac{(m_i^2-m_j^2)^2-q^2(m_i^2+m_j^2)}{q^4}\left(|A_t^L|^2+|A_t^R|^2\right),\nn \\
I_2^s(q^2) & \ = \ \dfrac{\lambda_q}{4 q^4}[|A_\perp^L|^2+|A_\parallel|^2+(L\to R)],\nonumber\\
I_2^c(q^2) & \ = \ -\dfrac{\lambda_q}{q^4}(|A_0^L|^2+|A_0^R|^2),
\end{align}
with the transversity amplitudes in terms of form factors and new Wilson coefficients are given as~\cite{Nebot:2007bc}
\bea
A_\perp^{L(R)} & \ = \ & N_{K^*}\sqrt{2} \lambda_B^{1/2}\left[(C_9^{\rm NP}\mp C_{10}^{\rm NP})\dfrac{V(q^2)}{M_B+M_{K^*}} \right], \nn \\
A_\parallel^{L(R)} & \ = \ &-N_{K^*}\sqrt{2}(M_B^2-M_{K^*}^2)\left[(C_9^{\rm NP}\mp C_{10}^{\rm NP})\dfrac{A_1(q^2)}{M_B-M_{K^*}}\right],\nonumber\\
A_0^{L(R)} & \ = \ & -\frac{N_{K^*}}{2 M_{K^*} \sqrt{q^2}}(C_9^{\rm NP}\mp C_{10}^{\rm NP}) \left( (M_B^2-M_{K^*}^2-q^2)(M_B+M_{K^*})A_1(q^2)-\frac{\lambda_B A_2 (q^2)}{M_B+M_{K^*}} \right),\nn \\ 
A_t^{L(R)} & \ = \ & - N_{K^*}\dfrac{\lambda_B^{1/2}}{\sqrt{q^2}}  (C_9^{\rm NP}\mp C_{10}^{\rm NP}) A_0(q^2),~
\eea
and 
\bea
&&\lambda_B=\lambda(M_B^2, M_{K^*}^2, q^2),~~\lambda_q=\lambda(m_i^2, m_j^2,q^2),\nn\\&&N_{K^*}(q^2) \ = \ V_{tb}V_{ts}^* \left[ \tau_{B_d}\frac{ \alpha_{\rm em}^2 G_F^2}{3\times 2^{10}\pi^5 M_B^3} \lambda_B^{1/2}\lambda_q^{1/2}\right]^{1/2}.
\eea

 After collecting the expressions for the LFV $B_{(s)}$ meson decay modes, we now proceed for the numerical estimation. We have taken the lifetime of $B_{(s)}$ meson, all the particles mass, the CKM matrix elments from PDG \cite{Zyla:2020zbs}\,,  $B\to K$ form factor from \cite{Ball:2004ye}, $B_{(s)} \to (K^*, \phi)$ form factor from \cite{Ball:2004rg,  Beneke:2004dp}. Using the upper limit of the allowed parameter space from Table \ref{Tab:conc}, the leptoquark mass as $M_{S_1}=1200$ GeV, and the vector-like fermion masses as $M_{\psi_q}=M_{N_1}=820$ GeV, $M_{N_2}=800$ GeV,  we have predicted the branching ratios of $B_s \to \mu^-\tau^+/\tau^-\mu^+$, $B^{+(0)} \to K^{+(0)} \mu^-\tau^+/\tau^-\mu^+$,  $B^{+(0)} \to K^{*+(0)} \mu^-\tau^+/\tau^-\mu^+$ and $B_s \to \phi \mu^-\tau^+/\tau^-\mu^+$ processes, which are tabulated in Table \ref{Tab:LFV}\,. 
\begin{table}[htb]
\begin{center}
\begin{tabular}{|c|c|c|c|c|}
\hline

Decay modes &Predicted branching ratios & Experimental Limits ($90\%$ CL)  \\ 
\hline
\hline
$B_s \to \mu^- \tau^+/\mu^+ \tau^-$~&~$5.373\times 10^{-8}$~&~~$<3.4\times 10^{-5}$~\cite{LHCb:2019ujz}\\

$B^+ \to K^+ \mu^- \tau^+/\mu^+ \tau^-$~&~$2.0\times 10^{-7}$~&~~$<2.8\times 10^{-5}/<4.5\times 10^{-5}$~\cite{BaBar:2012azg}\\

$\overline B^0 \to \overline K^0 \mu^- \tau^+/\mu^+ \tau^-$~&~$1.85\times 10^{-7}$~&~$\cdots$\\

$B^+ \to K^{* +} \mu^- \tau^+/\mu^+ \tau^-$~&~$2.971\times 10^{-7}$~&~$\cdots$\\

$\overline B^0 \to \overline K^{* 0} \mu^- \tau^+/\mu^+ \tau^-$~&~$2.742\times 10^{-7}$~~&~$\cdots$\\

$B_s \to \phi \mu^- \tau^+/\mu^+ \tau^-$~&~$3.592\times 10^{-7}$~&~$\cdots$\\

\hline
\end{tabular}
\end{center}
\caption{Predicted branching ratios of lepton flavor violating decay modes of $B_{(s)}$  meson. 
} \label{Tab:LFV}
\end{table} 
\begin{figure*}[htb]
\centering
\subfigure[~$B^+ \to K^+ \mu^-\tau^+$]{\includegraphics[width=0.48\textwidth]{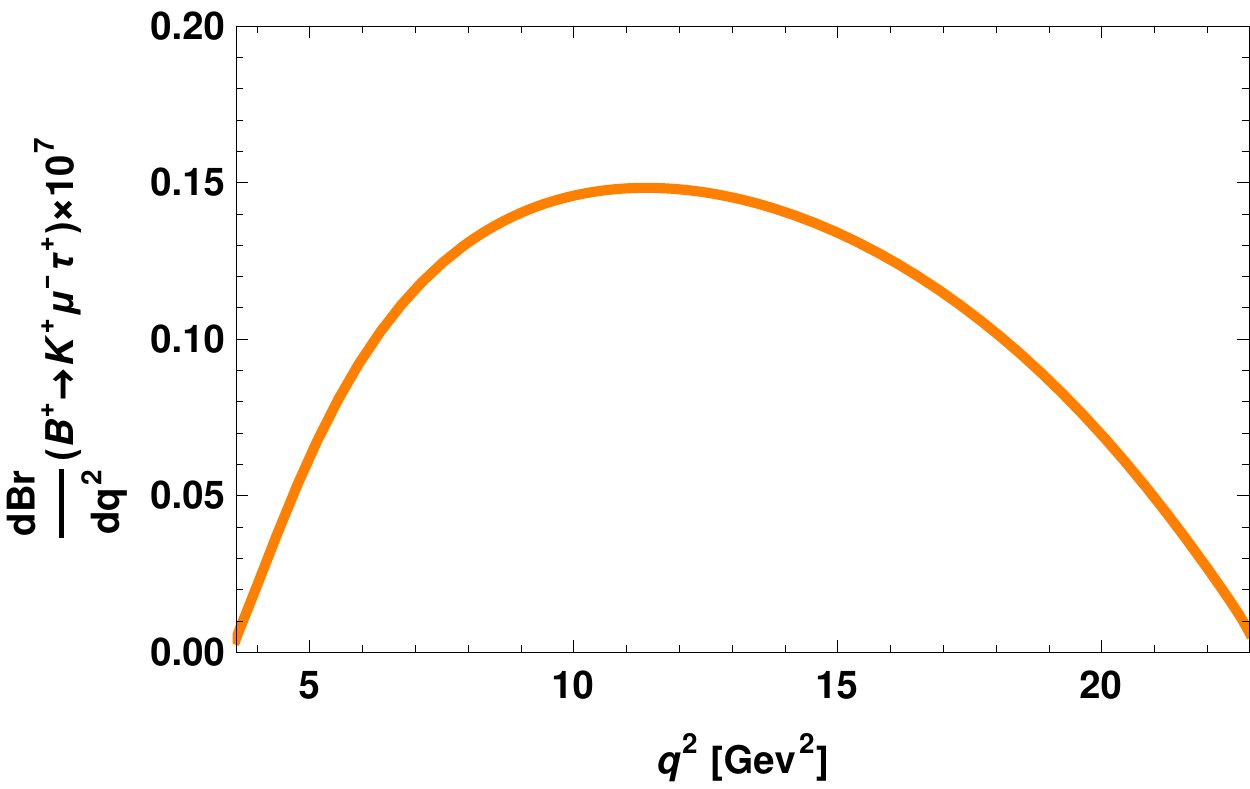}}
\quad
\subfigure[~$B^+ \to K^{*+} \mu^-\tau^+$]{\includegraphics[width=0.48\textwidth]{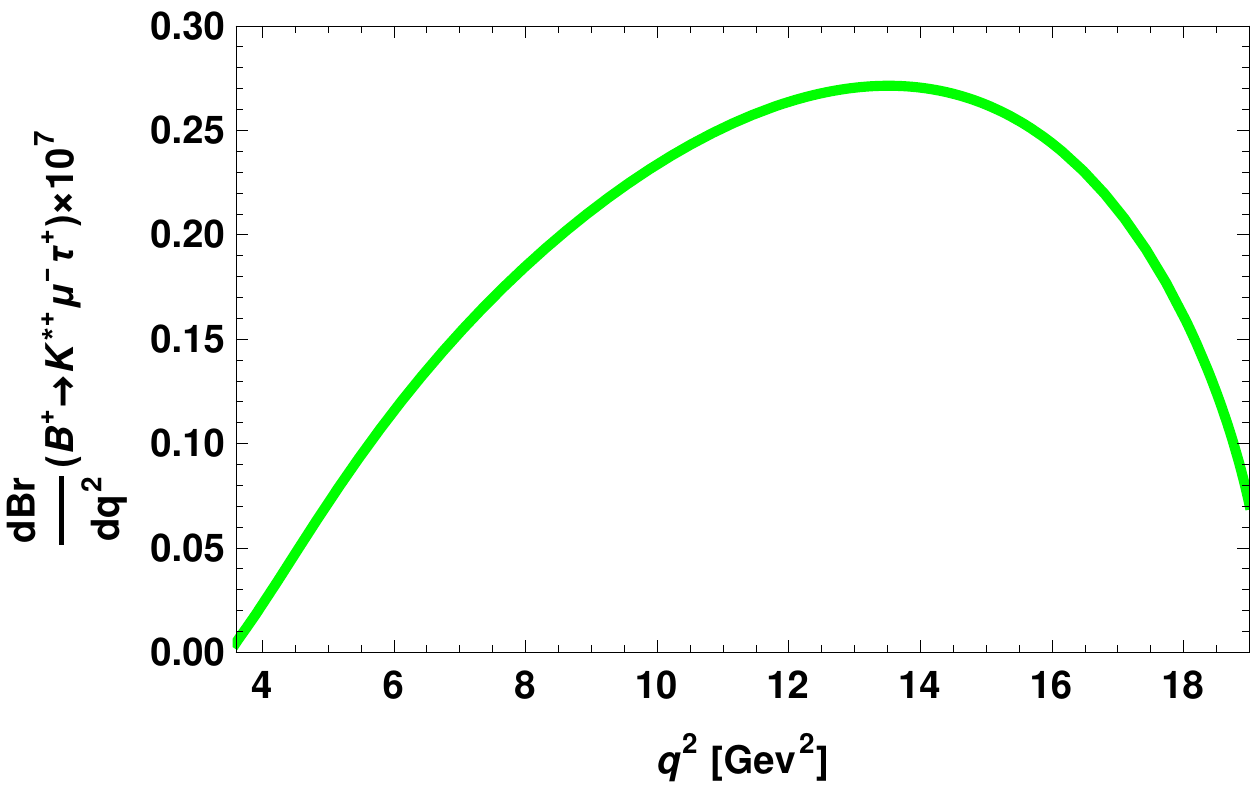}}
\quad
\subfigure[~$B_s \to \phi \mu^-\tau^+$]
{\includegraphics[width=0.5\textwidth]{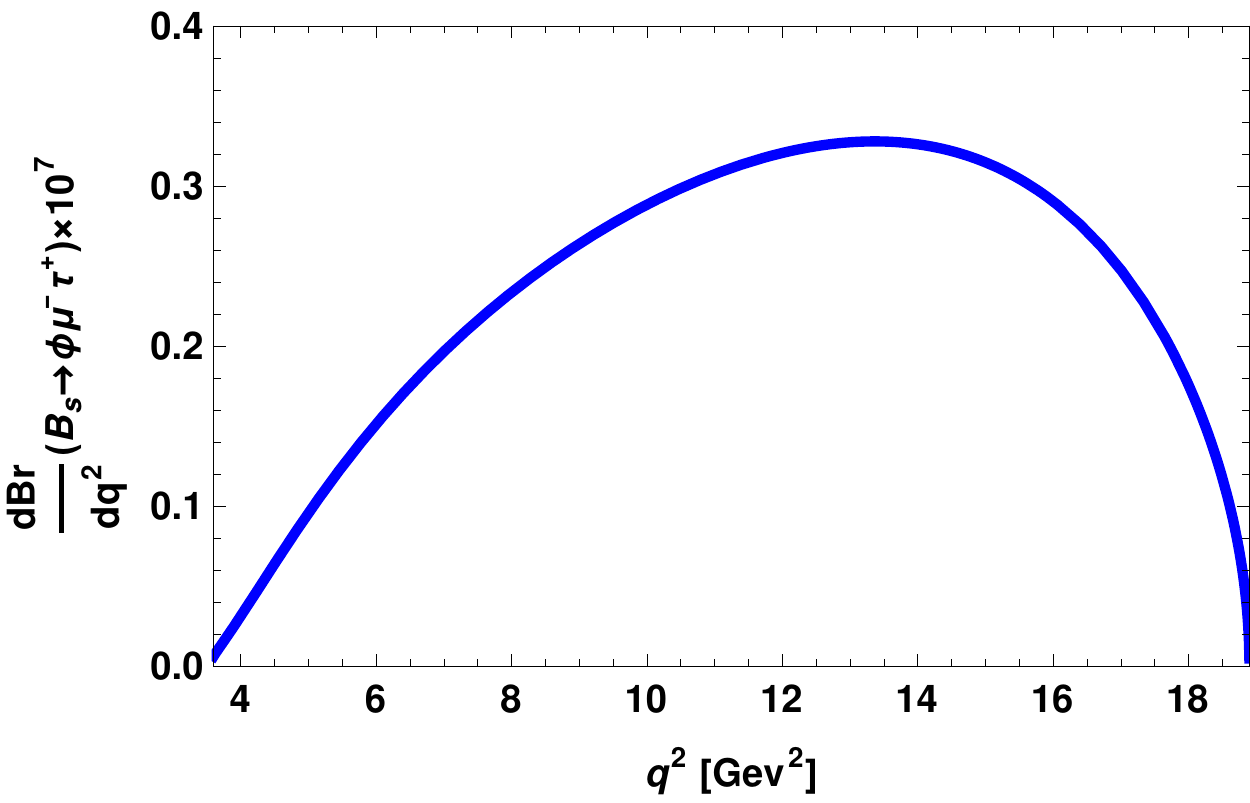}}
\caption{The $q^2$ variation of branching ratios of $B^+ \to K^+ \mu^- \tau^+$ (top-left panel), $B^+ \to K^{*+}\mu^- \tau^+$ (top-right panel) and $B_s \to \phi \mu^- \tau^+$ (bottom panel) processes. } \label{Fig:LFV}
\end{figure*}

 Though the experimental limits on most of these decay modes are  not yet available, from the table, one can notice that the branching ratios of the LFV $B$ decay modes  are within the reach of LHCb or $B$ factories.  There exist upper limit on  only $B_s\to \tau^\pm \mu^\mp$~\cite{LHCb:2019ujz} and  $B^+ \to K^+ \mu^-\tau^+ (\mu^+\tau^-) $~\cite{BaBar:2012azg} decay processes, for which our predicted branching ratios are well below the current 90\% CL experimental upper limits. Using the above mentioned allowed parameter space, our predictions on the branching ratio of $B_s\to \tau^\pm \mu^\mp$ process are
\bea
{\rm Br}(B_s\to \tau^\pm \mu^\mp)  =  {\rm Br}(B_s\to \tau^+ \mu^-)+{\rm Br}(B_s\to \tau^- \mu^+)  =  1.075\times 10^{-7}\,,
\eea 
within  the current  experimental limit at 90\% C.L.~\cite{LHCb:2019ujz}
\bea
{\rm Br}(B_s\to \tau^\pm \mu^\mp)|^{\rm Exp} \ < \ 3.4\times 10^{-5} \,.
\eea 
Fig. \ref{Fig:LFV} depicts the branching ratios of $B^+ \to K^+ \mu^-\tau^+$ (top-left panel), $B^+ \to K^{*+} \mu^-\tau^+$ (top-right panel) and $B_s \to \phi \mu^-\tau^+$ (bottom panel) LFV channels with respect to $q^2$.



\section{Comments on neutrino mass}
To realize neutrino mass, the model can be supplemented with three right-handed neutrinos ($i=3$) and allowing the Dirac interaction $y^i_\nu \overline{\ell_L}\tilde{H} \xi_{iR}$. Thus type-I seesaw provides $m_\nu \sim (y_\nu^{i} v)^2/M_{\xi_i}$. For a sample case, $M_{\xi_i} \sim 10$ TeV and $y^i_\nu \sim 10^{-5}$ gives sub-eV scale neutrino mass and also doesn't alter the phenomenological aspects discussed earlier in the paper.

\section{Concluding remarks}
The model is motivated to shed light on dark matter and also the existing anomalies in flavor sector, associated with $B$-meson. For the purpose, we extend standard model with vector-like fermions of quark and lepton type. Aided with a $(\overline{3},1,1/3)$ scalar leptoquark, we build a platform for new physics in flavor sector. An admixture of neutral vector-like lepton constitutes the relic density of the Universe through annihilation and co-annihilation channels mediated via scalar bosons, gauge bosons and vector-like fermions, leading to a freeze-out scenario. Apart from, the spin independent WIMP-nucleon cross section via $Z$-portal dictates the amount of mixing in neural vector-like leptons, and leptoquark-portal severely constrains the relevant Yukawa. Further, electroweak precision parameters constrain the mass splitting between the neutral vector-like components, to be above $\sim 7$ GeV.  In the presence of new vector-like fermions and scalar leptoquark, the rare $b \to s$ transitions occur through one loop box diagrams. By using the recent measurements on the branching ratios of $b \to sll(\nu_l\bar{\nu_l})$ and $b \to s \gamma$ processes such as $B_s \to ll$, $B\to K^{(*)} ll (\nu_l \bar \nu_l)$, $\bar B \to X_s \gamma$, and the lepton non-universality $R_{K^{(*)}}$ observables, we further constrained the new parameters like leptoquark couplings, vector-like fermions masses and couplings. We then estimate the branching ratios of rare (semi)leptonic $B_{(s)}$ decay modes which are found to be within the  reach of Belle II and LHCb experiments. 

\appendix
\acknowledgments 
 SS and RM would like to
acknowledge University of Hyderabad IoE project grant no. RC1-20-012. RM acknowledges the support from SERB,
Government of India, through grant No. EMR/2017/001448.

\section{Expressions of electroweak precision parameters}
\label{EWP_apdx}
The interaction of gauge interaction of vector-like fermions, written in a general form as
\begin{equation}
\mathcal{L} \supset \overline{\psi_a} (g_V \gamma^\mu + g_A \gamma^\mu \gamma_5)\psi_b V_\mu.
\end{equation}
Now, the electroweak parameters can be written in terms of $\Pi$ function as
\begin{equation}
\Pi(q^2) = \frac{N_c}{4\pi^2} \left[(g_V^2 + g_A^2) \tilde{\Pi}_{V\!+\! A}(q^2) + (g_V^2 - g_A^2)\tilde{\Pi}_{V\!+\! A}(q^2)\right],
\end{equation}
where, $N_c$ denotes the color charge of vector-like fermion. The expression for $\Pi$ functions and their derivatives at $q^{2}=0$ are given by \cite{Cynolter:2008ea}
\begin{eqnarray}
\tilde{\Pi}_{V\!+\! A}(0) & = & -\frac{(m_{a}^{2}+m_{b}^{2})}{2}
\left (\hbox{Div} +\ln\left(\frac{\mu^{2}}{m_{a}m_{b}}\right) \right ) -\frac{(m_{a}^{2}+m_{b}^{2} )}{4}  -\frac{\left(m_{a}^{4}+m_{b}^{4}\right)}{4\left(m_{a}^{2}-m_{b}^{2}\right)}
\ln\left(\frac{m_{b}^{2}}{m_{a}^{2}}\right),\nn\\
\tilde{\Pi}_{V\!-\! A}(0) & = & m_{a}m_{b}\left(\hbox{Div}+\ln\left(\frac{\mu^{2}}{m_{a}m_{b}}\right) +\frac{\left(m_{a}^{2}+m_{b}^{2}\right)}{2\left(m_{a}^{2}-m_{b}^{2}\right)}\ln\left(\frac{m_{b}^{2}}{m_{a}^{2}}\right)+1\right),
\end{eqnarray}
and the derivatives 
\begin{eqnarray}
\tilde{\Pi}'_{V\!+\! A}(0) & \!\!=\!\!\! & 
 \frac{1}{3}\hbox{Div} +\ln\left(\frac{\mu^{2}}{m_{a}m_{b}}\right)  \!+\!\frac{(m_{a}^{4}-8m_{a}^{2}m_{b}^{2}+m_{b}^{4})}{9\left(m_{a}^{2}-m_{b}^{2}\right)^{2}} +\frac{\left(m_{a}^{2}+m_{b}^{2}\right)\left(m_{a}^{4}-4m_{a}^{2}m_{b}^{2}+m_{b}^{4}\right)}{6\left(m_{a}^{2}-m_{b}^{2}\right)^{3}}\ln\left(\frac{m_{b}^{2}}{m_{a}^{2}}\right), \nonumber\\
\tilde{\Pi}'_{V\!-\! A}(0) & = & \frac{m_{a}m_{b}\left(m_{a}^{2}+m_{b}^{2}\right)}{2\left(m_{a}^{2}-m_{b}^{2}\right)^2}+\frac{m_{a}^{3}m_{b}^{3}}{\left(m_{a}^{2}-m_{b}^{2}\right)^{3}}\ln\left(\frac{m_{b}^{2}}{m_{a}^{2}}\right).
\end{eqnarray}
and \begin{eqnarray}
\tilde{\Pi}''_{V\!+\! A}(0) & = & \frac{\left(m_{a}^{2}+m_{b}^{2}\right)\left(m_{a}^{4}-8m_{a}^{2}m_{b}^{2}+m_{b}^{4}\right)}{4\left(m_{a}^{2}-m_{b}^{2}\right)^{4}}-\frac{3m_{a}^{4}m_{b}^{4}}{\left(m_{a}^{2}-m_{b}^{2}\right)^{5}}\ln\left(\frac{m_{b}^{2}}{m_{a}^{2}}\right),\nonumber\\
\tilde{\Pi}''_{V\!-\! A}(0) & = & \frac{m_{a}m_{b}\left(m_{a}^{4}+10m_{a}^{2}m_{b}^{2}+m_{b}^{4}\right)}{3\left(m_{a}^{2}-m_{b}^{2}\right)^{4}}+\frac{2\left(m_{a}^{2}+m_{b}^{2}\right)m_{a}^{3}m_{b}^{3}}{2\left(m_{a}^{2}-m_{b}^{2}\right)^{5}}\ln\left(\frac{m_{b}^{2}}{m_{a}^{2}}\! \right).
\end{eqnarray}
For identical masses i.e., $m_{a}=m_{b}$, the above functions take the form
\begin{eqnarray}
&&\tilde{\Pi}{}_{V\!+\! A}(0)=-m_{a}^{2} \hbox{Div}-m_{a}^{2} \ln \left(\frac{\mu^{2}}{m_{a}^2}\right), \nn\\
&&\tilde{\Pi}{}_{V\!-\! A}(0)=m_{a}^{2}\hbox{Div} +m_{a}^{2} \ln \left(\frac{\mu^{2}}{m_{a}^2}\right),\nn\\
&&\tilde{\Pi}'_{V\!+\! A}(0)=\frac{1}{3}\hbox{Div}+\frac{1}{3} \ln\left(\frac{\mu^{2}}{m_{a}^2}\right) -\frac{1}{6},\nn\\
&&\tilde{\Pi}'_{V\!-\! A}(0)=\frac{1}{6},\nn\\
&&\tilde{\Pi}''_{V\!+\! A}(0)=\frac{1}{10m_{a}^{2}}, \nn\\
&&\tilde{\Pi}''_{V\!-\! A}(0)=\frac{1}{30m_{a}^{2}}.
\end{eqnarray}
\section{$J_{1,2}^{s(c)}$ functions of $B \to K^* l^+l^-$  process}
The $J_{1,2}^{s(c)}$ functions required to compute the decay rate of $B \to K^* ll$ processes are given by \cite{Bobeth:2008ij}
\bea
 J^s_1 &=& \frac{\left(2+\beta ^2_l\right)}{4}\Bigg[|A_\perp ^L|^2 + |A_\parallel ^L|^2 + \left(L\rightarrow R\right)\Bigg] + \frac{4m^2_l}{q^2} 
{\rm Re}\left(A_\perp ^L A_\perp ^{R^*} + A_\parallel ^L A_\parallel ^{R^*}\right),~~ \\
  J^c_1 &=& |A_0^L|^2 + |A_0^R|^2 +\frac{4m^2_l}{q^2}\Bigg[|A_t|^2 + 2{\rm Re}\left(A_0^L A_0^{R^*}\right)\Bigg], \\
  J^s_2 &=& \frac{\beta^2_l}{4}\left[|A_\perp ^L|^2 + |A_\parallel ^L|^2 + \left(L\rightarrow R\right)\right], \\
  J^c_2 &=& -\beta^2_l\left[|A_0^L|^2 +\left(L\rightarrow R\right)\right], \\
 \eea
 with
 \bea
 A_i A_j^* = A_{i}^{ L}\left(q^2\right) A^{* L}_{j}\left(q^2\right) + A_{i}^{ R}\left(q^2\right) A^{* R}_{j}\left(q^2\right) 
\hspace{1cm} \left(i,j = 0, \parallel, \perp\right),
 \eea
 in shorthand notation. 
\section{Loop functions of muon anomalous magnetic moment}
The loop functions involved in the scalar leptoquark contribution to muon anomalous magnetic moment  are  given as \cite{Lavoura:2003xp}
\bea
&&f_1(x_F)= m_1 \left(c_1+\frac{3}{2}d_1 \right),    ~~~~~~\bar f_1(x_F)= m_1 \left(-\bar c_1+\frac{3}{2}\bar d_1 \right),  \\
&&f_2(x_F)= m_2 \left(c_2+\frac{3}{2}d_2 \right),    ~~~~~~\bar f_2(x_F)= m_2 \left(-\bar c_2+\frac{3}{2}\bar d_2 \right),    
\eea
where
\bea
&&c=\frac{x_F-3}{4(x_F-1)^2}+\frac{\log x_F}{(x_F-1)^3},~~~d=\frac{-2x_F^2+7x_F-11}{18(x_F-1)^3}+\frac{\log x_F}{3(x_F-1)^4},\\
&& \bar c =\frac{3x_F-1}{4(x_F-1)^2}+\frac{x_F^2\log x_F}{2(x_F-1)^3}, ~~~ \bar d= \frac{11x_F^2-7x_F+2}{18(x_F-1)^3}-\frac{x_F^3\log x_F}{3(x_F-1)^4}. 
\eea



\bibliography{vectorlike}

\end{document}